\documentclass[sigconf]{acmart}

\usepackage{booktabs}
\AtBeginDocument{%
  }

\usepackage{multirow}
\acmConference[Conference acronym 'XX]{Make sure to enter the correct
  conference title from your rights confirmation email}{June 03--05,
  2018}{Woodstock, NY}
\acmISBN{978-1-4503-XXXX-X/2018/06}

\begin{document}

\title{HASTE: Hierarchical and Attribute-aware Supervision for Generative Recommendation}

\author{Tripti Shukla}
\email{tripti.shukla@glance.com}
\affiliation{%
  \institution{Glance}
  \city{Bengaluru}
  \country{India}
}

\author{Shwetha Somasundaram}
\email{shwetha.s@glance.com}
\affiliation{%
  \institution{Glance}
  \city{Bengaluru}
  \country{India}
}

\author{Akshat Gupta}
\email{akshat.gupta@glance.com}
\affiliation{%
  \institution{Glance}
  \city{Bengaluru}
  \country{India}
}

\author{Rohit Garg}
\email{rohit.garg@glance.com}
\affiliation{%
  \institution{Glance}
  \city{Bengaluru}
  \country{India}
}

\author{Paul Duff}
\email{paul.duff@glance.com}
\affiliation{%
  \institution{Glance}
  \city{Vilnius}
  \country{Lithuania}
}

\author{Ian Anderson}
\email{ian.anderson@glance.com}
\affiliation{%
  \institution{Glance}
  \city{London}
  \country{United Kingdom}
}

\renewcommand{\shortauthors}{Shukla et al.}

\begin{abstract}
  Generative recommendation models have shown promising results by representing items as semantic IDs and generating these IDs autoregressively. We introduce an approach showing that explicitly incorporating hierarchical and attribute-level supervision into generative recommendation can simultaneously improve training efficiency and long-tail recommendation performance. Existing approaches typically optimize semantic ID prediction with a standard causal cross-entropy objective, which does not explicitly exploit the hierarchical structure of semantic IDs and can require substantial training to learn effective item representations. Moreover, data-sparse tail items often suffer from poor representations due to limited supervision, resulting in weaker long-tail recommendation performance. We address these challenges with HASTE (\textbf{H}ierarchical and \textbf{A}ttribute-aware \textbf{S}upervision for genera\textbf{T}ive r\textbf{E}commendation), a unified training framework that introduces hierarchical and attribute-specific supervision. HASTE's Hierarchical Loss explicitly exploits the multi-level structure of semantic IDs, providing supervision at each level of the item hierarchy and substantially accelerating training convergence. Its Attribute-Specific Losses encourage the model to preserve fine-grained item attributes, leveraging the language model's inherent semantic knowledge to improve representations and recommendations for tail items. We extensively evaluate HASTE on a fashion recommendation benchmark and across multiple domains of the Amazon dataset. HASTE improves recommendation performance across datasets and item-frequency regimes. Notably, the Hierarchical Loss achieves approximately 2× faster convergence than the standard causal cross-entropy objective, substantially improving training efficiency while maintaining strong recommendation quality.
\end{abstract}




\maketitle

\section{Introduction}

Large language models (LLMs) have triggered a paradigm shift across modalities, replacing task-specific pipelines with a single model that is pre-trained once and then adapted through instructions or lightweight fine-tuning~\cite{brown2020gpt3,touvron2023llama}. Recommender
systems are undergoing an analogous shift. Industrial recommenders are conventionally built as cascades, commonly described as matching, pre-ranking,
ranking, and re-ranking~\cite{ma2021pre,song2022rethinking}, whose stages are developed and trained as separate models with their own objectives: YouTube's
two-stage system, for instance, trains candidate generation as an extreme
multiclass classification problem and ranking under a watch-time-weighted
logistic objective~\cite{covington2016youtube}, while
pre-ranking~\cite{wang2020cold} and re-ranking~\cite{pei2019prm} each introduce
objectives of their own. Generative recommendation instead casts the entire
retrieval problem as a single sequence-to-sequence task: given a user's
interaction history, the model \emph{generates} the identifier of the next item
directly. ~\cite{geng2022p5}
TIGER~\cite{rajput2023recommendersystemsgenerativeretrieval} established the
form this now takes, replacing arbitrary item IDs with residual-quantized semantic IDs.
 Subsequent work has shown that ranking and
retrieval can be unified in a single generative
backbone~\cite{zhai2024actionsspeaklouderwords,deng2025onerec}, whose quality
scales as a power law of training
compute~\cite{zhai2024actionsspeaklouderwords}.

A parallel line of work marries generative recommendation with LLM backbones in
order to exploit the world knowledge and semantic priors that recommendation
data alone cannot supply~\cite{zheng2024lcrec,he2026plum}. The bridge between the two is the \emph{semantic ID} (SID): rather than
assigning each item an atomic, randomly initialised embedding, an item is
represented as a short sequence of discrete codes derived from its content
representation. Semantically similar items share prefixes, so the identifier
space itself carries item semantics, and the model can generalise to items with
little or no interaction
history~\cite{rajput2023recommendersystemsgenerativeretrieval}. Semantic IDs are
typically produced by residual quantization---either
RQ-VAE~\cite{lee2022rqvae,rajput2023recommendersystemsgenerativeretrieval} or
the training-free residual $k$-means of
Quantitative Alignment Multi-Modal Recommendation~\cite{luo2024qarm}---and are
therefore \emph{hierarchical by
construction}. Crucially, this makes the levels
\emph{asymmetric}: because every later code quantizes the residual left by the
earlier ones, an error at a coarse level redefines the target of every level
below it, whereas an error at the finest level perturbs only a local refinement.

\begin{figure*}[t]
  \centering
 \includegraphics[width=\linewidth,height=12cm]{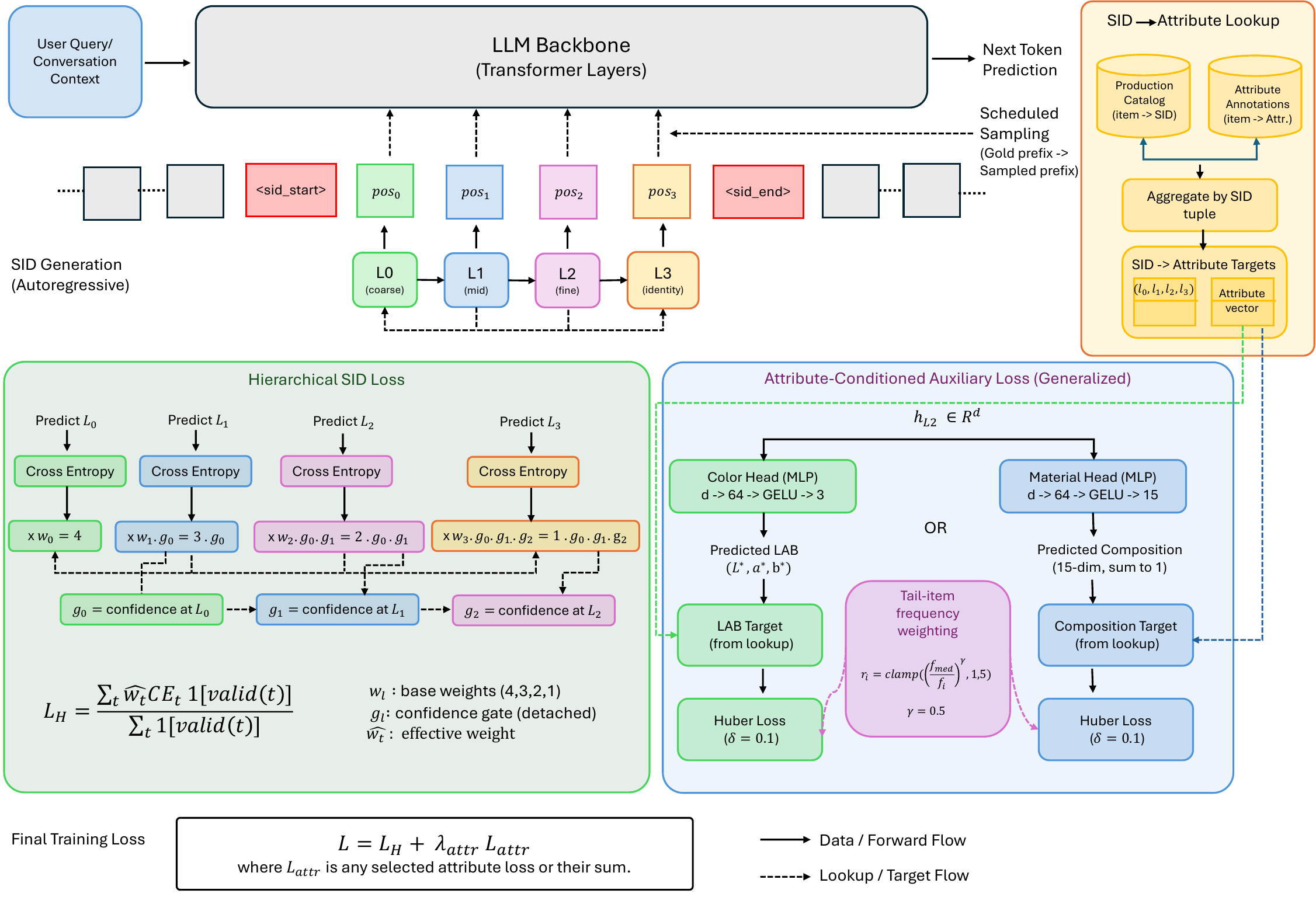}
  \caption{Overview of the proposed \textit{HASTE} training framework. The LLM autoregressively predicts hierarchical SIDs $L_0,L_1,L_2,L_3$ with confidence-gated loss $\mathcal{L}_H$, where $w_i$ and $g_i$ denote level weights and detached confidence gates. The LLM hidden state $h_{L2}\in\mathbb{R}^d$ feeds a selected attribute-specific MLP, with $\mathcal{L}_{\mathrm{attr}}$ weighted by $\lambda_{\mathrm{attr}}$. The SID tuple $(l_0,l_1,l_2,l_3)$ indexes its corresponding attribute vector.}
  \label{fig:overview}
\end{figure*}

Despite this structure, essentially all existing generative recommenders
supervise semantic-ID generation with a flat causal cross-entropy (CE) loss over
the extended vocabulary. The consequence is not merely that the model must \emph{rediscover} the coarse-to-fine organisation residual quantization already encoded---it is that flat CE loss actively \emph{erodes} it. We
show in Section~\ref{sec:motivation} that fine-tuning under flat CE cuts the
cosine coupling between adjacent SID levels by $35\%$ relative to initialization
(from $0.55$ to $0.36$), while a level-aware objective retains it at $0.50$, a
$9\%$ reduction, and induces $2$--$4\times$ less embedding drift from the shared
initialization.

A second, orthogonal limitation is that the supervision an item receives is
proportional to how often it appears as a training target. Head items accumulate
enough gradient to acquire useful representations; data-sparse tail items do
not. This
matters because semantic IDs are quantized from \emph{holistic} item
embeddings, so they organise the catalogue by overall similarity rather than by
any single attribute: a property a user actually asks for (color, material,
fit, pattern) is smeared across levels rather than localised. Measuring colour
directly, the coarsest code alone yields only $37\%$ mean colour purity while
conditioning on the first two codes raises it to $83\%$, and a probing analysis
shows the information is present but not linearly organised---a linear probe
underperforms the catalogue-mean baseline at every SID position, whereas a small
MLP probe reaches $\Delta E = 38.5$ against a baseline of $56.3$. The signal is
latent in the representation and unexploited by the generation objective. Yet the
LLM backbone already \emph{knows} what a ``ribbed burgundy wool cardigan'' is:
the interaction log is the bottleneck, not the backbone's semantic knowledge.

To address these limitations, we propose \textbf{HASTE} (\textbf{H}ierarchical and \textbf{A}ttribute-aware
\textbf{S}upervision for genera\textbf{T}ive r\textbf{E}commendation), a
fine-tuning framework that introduce two complementary objectives. The
\textbf{Hierarchical Loss} replaces flat causal CE with level-aware supervision:
coarser levels carry larger base weights, and each level's weight is gated by
the model's \emph{detached} confidence in the upstream codes it is conditioned
on, so fine-grained supervision is amplified only once the coarse decisions it
depends on are actually resolved. Because confidence estimates are unreliable early in training, the objective anneals from purely static hierarchical weighting to fully confidence-gated weighting. Scheduled sampling~\cite{bengio2015scheduledsampling} further perturbs the upstream prefix, exposing the model to the upstream errors it is likely to encounter during decoding rather than only to clean, teacher-forced trajectories. The
\textbf{Attribute-Specific Losses} attach lightweight MLP regression heads to
the hidden state already used to predict the fine-grained SID code, and
supervise them against catalogue-derived attribute targets. A
frequency-aware weighting further concentrates the signal on infrequent items.
We instantiate the framework for two complementary attributes---perceptual
(CIELAB colour~\cite{CIE1976Lab}) and compositional (material composition)---and
find that the gains land where interaction supervision is weakest: mean top-1
colour error drops from $54.5$ to $50.2$ on tail colours and $33.5$ to $29.0$ on
torso colours, while head colours are unchanged ($12.5$ to $12.6$).

Our contributions are as follows:
\begin{itemize}
  \item We propose a \textbf{Hierarchical Loss} that supervises semantic-ID
  generation level-by-level, combining decreasing level weights, detached
  confidence gating on upstream codes, an annealed transition from static to
  conditional weighting, and scheduled sampling of the upstream prefix to close
  the train--inference exposure gap. It exploits the hierarchy explicitly rather
  than leaving it to be rediscovered, preserves the inherited structure, and
  yields $\sim$$2\times$ faster convergence than causal CE at equal or better
  recommendation quality.
  
  \item We propose \textbf{Attribute-Specific Losses} that inject fine-grained
  item-attribute supervision into the generative recommender through auxiliary
  heads on the fine-grained SID prediction state.  (Section~\ref{sec:attribute_loss}).
  

  \item We evaluate HASTE on three Amazon Review~\cite{hou2024bridging} subsets under the standard P5 5-core
  protocol~\cite{geng2022p5}, on Fashion-140k, and on an internal production-scale fashion catalog, against sequential, generative, and LLM-based baselines. HASTE
  gives prominent gains across datasets and item-frequency regimes, with the
  largest improvements on the tail items.
\end{itemize}

\section{Related Works}

\noindent\textbf{Sequential and Generative Recommendation.}
Sequential recommenders predict the next item from an ordered interaction
history. GRU4Rec~\cite{hidasi2016gru4rec} models sessions with recurrence, while
SASRec~\cite{kang2018selfattentivesequentialrecommendation} and
BERT4Rec~\cite{sun2019bert4recsequentialrecommendationbidirectional} use
unidirectional and bidirectional self-attention respectively. All three represent
an item as an atomic embedding in a table that grows with the catalogue, so a
rare item's representation is learned almost entirely from its few occurrences,
and retrieval needs a nearest-neighbour index maintained separately from the
model. Generative retrieval removes the index: the model decodes the target's identifier
token by token~\cite{tang2023gensurvey}. GENRE~\cite{decao2021genre} first cast
retrieval this way, generating entity names autoregressively;
DSI~\cite{tay2022dsi} replaced names with structured semantic identifiers
obtained by hierarchical clustering; and NCI~\cite{wang2022nci} added a
prefix-aware decoder. P5~\cite{geng2022p5} first brought this seq2seq formulation to recommendation,
and TIGER~\cite{rajput2023recommendersystemsgenerativeretrieval} specialised it
to semantic IDs, quantizing content embeddings with
RQ-VAE~\cite{lee2022rqvae} and decoding the next item's codes
from the history; because those codes are derived from content and shared
prefix-wise, a rare item is no longer represented in isolation.
HSTU~\cite{zhai2024actionsspeaklouderwords} showed the paradigm scales as a power
law of training compute, and OneRec~\cite{deng2025onerec} and
OneRec-Think~\cite{liu2025onerecthinkintextreasoninggenerative} deploy it at
industrial scale.
In addition to the generation framework, item tokenization has also attracted
considerable attention. Recforest~\cite{feng2022recforest} applies hierarchical
$k$-means to item text embeddings and uses the resulting cluster path as the
identifier, while QARM~\cite{luo2024qarm} replaces the learned quantizer with a
training-free residual $k$-means. Furthermore, several studies incorporate
collaborative signals into the tokenization process:
MMGRec~\cite{liu2024mmgrec} performs quantization over a user--item interaction
graph, EAGER~\cite{wang2024eager} decodes behavioural and semantic tokens in two
parallel streams, and LETTER~\cite{wang2024letter} augments a learnable tokenizer
with collaborative and diversity regularisers. 

\noindent\textbf{LLMs for Recommendation.}
Chat-Rec~\cite{gao2023chatrec} prompts a general-purpose model with textual
interaction histories, and GPT4Rec~\cite{li2023gpt4rec} fine-tunes one to generate
item titles; such models rank competently only over small candidate pools and
remain sensitive to interaction order, item position, and
popularity~\cite{hou2024zeroshot}. Recommenders built on language-model backbones
tune on recommendation data instead: P5~\cite{geng2022p5} unifies several tasks as
text-to-text but with arbitrary numeric item IDs,
LC-Rec~\cite{zheng2024lcrec} aligns semantic-ID tokens with language through
auxiliary alignment tasks, and
IDGenRec~\cite{tan2024idgenrecllmrecsysalignmenttextual} generates textual
identifiers instead of quantized codes. This work follows the semantic-ID line and
investigates the \emph{training objective} itself, given a fixed architecture.
Whereas LC-Rec introduces additional alignment \emph{tasks}, HASTE reshapes the
loss on the recommendation task and derives attribute targets from the catalogue,
requiring no additional task data or prompt engineering.

\noindent\textbf{Structured and Auxiliary Supervision.}
Label structure is well exploited outside recommendation. Hierarchical
softmax~\cite{morin2005hsoftmax} factorises the prediction over a tree,
coarse-to-fine curricula~\cite{stretcu2021coarsetofine} order the levels in time,
and hierarchy-aware penalties~\cite{bertinetto2020hierarchy} charge more for
distant errors than for sibling errors. Auxiliary objectives likewise densify
supervision: deeply-supervised nets~\cite{lee2015deeply} attach classifiers at
intermediate layers, MMoE~\cite{ma2018mmoe} shares experts across tasks in the
multi-task architectures common to recommenders, and
CLCRec~\cite{wei2021contrastive} uses a contrastive objective to preserve
collaborative signal in the content representations of cold-start items.
Scheduled sampling~\cite{bengio2015scheduledsampling} addresses the exposure bias
of autoregressive decoders, and applies to semantic IDs, whose
levels are defined relative to their predecessors. Semantic IDs are a residual hierarchy over an autoregressively decoded sequence, so
all three ideas apply directly. HASTE combines level-conditioned supervision
over SID codes with attribute heads that recycle the backbone's semantic knowledge
into tail-item representations.

\section{Preliminaries}
\label{sec:preliminaries}
\subsection{Generative Recommendation with Semantic IDs}
\label{sec:generative_recommendation}
Generative recommendation~\cite{rajput2023recommendersystemsgenerativeretrieval} formulates item retrieval as a sequence generation problem, where a language model generates the identifier of a recommended item conditioned on the user's interaction history. Recent approaches represent items using \emph{semantic IDs} (SIDs), replacing independent item identifiers with sequences of discrete tokens derived from the item's semantic representation. For an item $i$, its semantic ID is represented as
\begin{equation}
\mathrm{SID}_i =
\langle\mathrm{sid\_start}\rangle
\;s_{i,0}\;s_{i,1}\;s_{i,2}\;s_{i,3}\;
\langle\mathrm{sid\_end}\rangle,
\end{equation}
where $s_{i,0},\ldots,s_{i,3}$ denote the semantic-ID codes and the start and end tokens delimit the SID sequence. 

Given a user interaction context $x$, the language model autoregressively generates the semantic-ID codes. The probability of generating the SID can be factorized as
\begin{equation}
P(\mathrm{SID}_i \mid x)
=
\prod_{l=0}^{3}
P_{\theta}
\left(
s_{i,l} \mid x,s_{i,<l}
\right),
\end{equation}
where $s_{i,<l}$ denotes the semantic-ID codes generated before level $l$. 
The generated SID is subsequently mapped to the corresponding item in the catalog. Unlike conventional item-ID-based recommendation, semantic IDs allow the model to operate over a structured representation of the item space. The effectiveness of this formulation therefore depends not only on the semantic structure encoded by the SIDs, but also on how effectively the training objective exploits this structure during generation.

\subsection{Residual Quantization for Semantic IDs}
\label{sec:rqvae}
Semantic IDs can be obtained using residual quantization~\cite{zeghidour2021soundstreamendtoendneuralaudio}, which represents an item embedding through a sequence of progressively refined discrete codes. Given an item representation $z$, residual quantization first selects a code from a coarse codebook and subsequently quantizes the residual using additional codebooks. The resulting representation can be expressed as
\begin{equation}
z \approx
e_{s_{i,0}}^{(0)}
+
e_{s_{i,1}}^{(1)}
+
e_{s_{i,2}}^{(2)}
+
e_{s_{i,3}}^{(3)},
\end{equation}
where $e_{s_{i,l}}^{(l)}$ denotes the code selected from the $l$-th quantization codebook.\footnote{We treat all codebooks as comparably sized and expressive; this is the common setup in RQ-VAE-based SID construction and is not a requirement of residual quantization itself.}

This construction induces a natural coarse-to-fine hierarchy among the semantic-ID codes: because $s_{i,l}$ is chosen to quantize the residual $z - \sum_{k<l} e_{s_{i,k}}^{(k)}$ left after the first $l$ levels, level $l$'s code is defined only relative to the codes chosen at levels $0,\ldots,l-1$. Consequently, earlier codes capture broader semantic characteristics of the item, while subsequent codes encode residual information that progressively refines the representation, and the semantic meaning of a later code is conditioned on the codes that precede it. A key implication, which we return to in Section~\ref{sec:sid_prediction}, is that an incorrect code at an early level changes the residual every later level is quantizing against — so errors at coarse levels are structurally more consequential than errors at fine levels.


\subsection{Semantic-ID Prediction with Causal Language Models.}
\label{sec:sid_prediction}
Under the standard next-token prediction objective, each semantic-ID code is supervised using causal cross-entropy, summed over the training corpus $\mathcal{D}$:
\begin{equation}
\mathcal{L}_{\mathrm{CE}}
=
-\sum_{i \in \mathcal{D}}\sum_{l=0}^{3}
\log
P_{\theta}
\left(
s_{i,l}
\mid
x,s_{i,<l}
\right).
\end{equation}

The objective provides supervision for each semantic-ID prediction while preserving the autoregressive generation process. However, it weights the loss at every level equally: a mistake on the coarse code $s_{i,0}$ and a mistake on the fine code $s_{i,3}$ contribute the same magnitude to $\mathcal{L}_{\mathrm{CE}}$. This uniform weighting is at odds with the coarse-to-fine structure established in Section~\ref{sec:rqvae} since an error at an early level corrupts the semantic context for every level after it, while an error at a late level only affects that level's refinement, the two errors are not equally costly. 

This distinction is important in semantic-ID-based recommendation: the training objective does not reflect the hierarchical relationship induced by residual quantization, even though the model's inputs and outputs do. We build on this observation by introducing hierarchy-aware supervision in HASTE.

\section{Motivation}
\label{sec:motivation}

\begin{figure}[h]
  \centering
 \includegraphics[width=\linewidth]{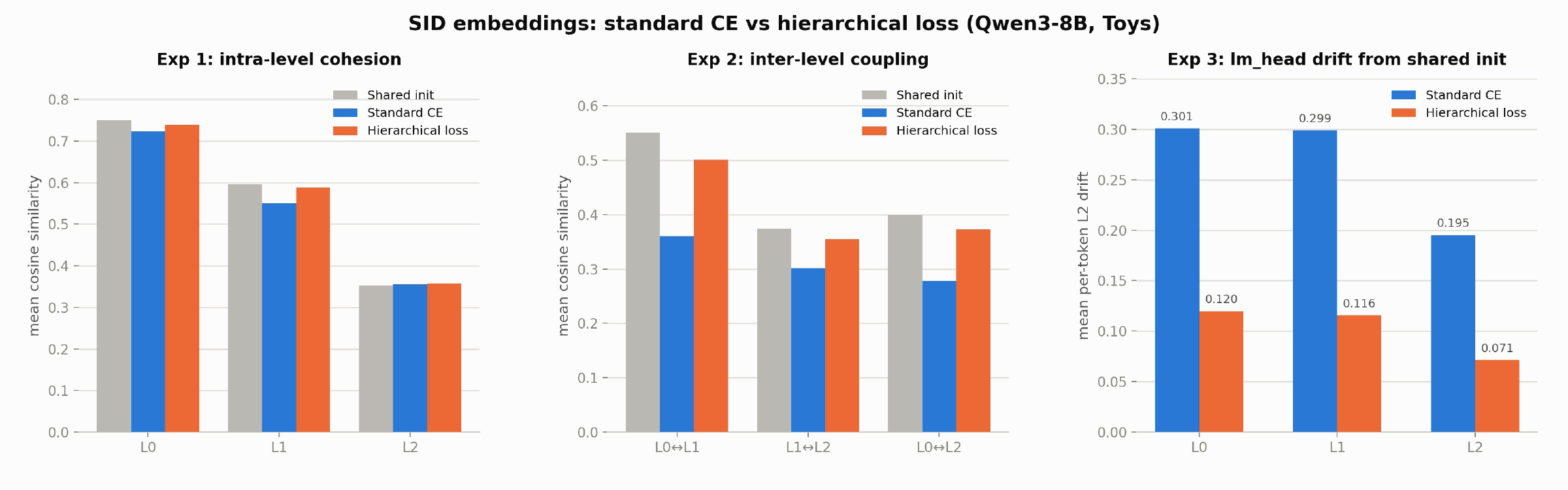}
  \caption{Hierarchical structure of semantic-ID representations under different fine-tuning objectives. (a) Within-level cohesion, (b) inter-level coupling, and (c) embedding drift from the shared initialization. Hierarchical Loss better preserves the structure inherited from residual quantization than standard causal cross-entropy.}
  \label{fig:hier_embed_anal}
  \Description{bar charts comparing std CE \& hierarchical loss.}
  
\end{figure}

\noindent \textbf{Preserving the Hierarchical Structure of Semantic IDs.} Semantic IDs (SIDs) generated by residual quantization encode items through a coarse-to-fine hierarchy, where earlier codes capture broad semantics and later codes progressively refine the representation. However, standard causal cross-entropy treats SID tokens as independent vocabulary predictions and does not explicitly model dependencies across hierarchy levels. This raises the question of whether the structure encoded by RQ-VAE is preserved during language-model fine-tuning.


We compare models trained with standard causal cross-entropy and our Hierarchical Loss, both initialized from the same vocabulary-aligned checkpoint, obtained after extending the language model vocabulary with the SID tokens and training their embeddings to align with the pretrained representation space, and subsequently subjected to full-parameter supervised fine-tuning. As shown in Figure~\ref{fig:hier_embed_anal}, we analyze the SID representations along three complementary dimensions: within-level cohesion, inter-level coupling, and embedding drift from the shared initialization. Standard CE substantially weakens the hierarchical structure: the average cosine similarity between L0 and L1 output embeddings decreases from $0.55$ at initialization to $0.36$ after fine-tuning, a $35\%$ reduction. In contrast, Hierarchical Loss preserves this relationship at $0.50$, corresponding to only a $9\%$ reduction. We observe similar trends across other level relationships, while the coarse L0 representations also retain greater within-level cohesion under hierarchical supervision.

Moreover, relative to the shared initialization, Hierarchical Loss induces 2$-$4$\times$ less embedding drift than standard CE across all SID levels and embedding tables. We measure drift as the mean per-token Euclidean distance $\lVert v_{\text{finetuned}} - v_{\text{init}} \rVert_2$ between each SID token's row vector after fine-tuning and its value at the shared initialization. These results suggest that a flat token-level objective can weaken the coarse-to-fine structure inherited from residual quantization. This motivates our \textbf{Hierarchical Loss}, which explicitly incorporates dependencies across SID levels, encouraging the model to learn coarse semantics before progressively refining them into fine-grained item representations.

\noindent \textbf{Improving Tail-Item Representations with Attribute Supervision.} While hierarchical supervision addresses the structure of semantic IDs, it does not explicitly ensure that the learned representations preserve individual item attributes. This is particularly important for generative recommendation, where users may express preferences for specific attributes such as color, material, fit, or pattern. Semantic IDs are derived from holistic item representations and therefore organize products according to overall semantic similarity rather than any single attribute. Consequently, an attribute relevant to a user's query may be distributed across multiple levels of the semantic-ID hierarchy and need not be explicitly preserved by the standard next-token prediction objective.

\begin{figure}[h]
  \centering
 \includegraphics[width=\linewidth]{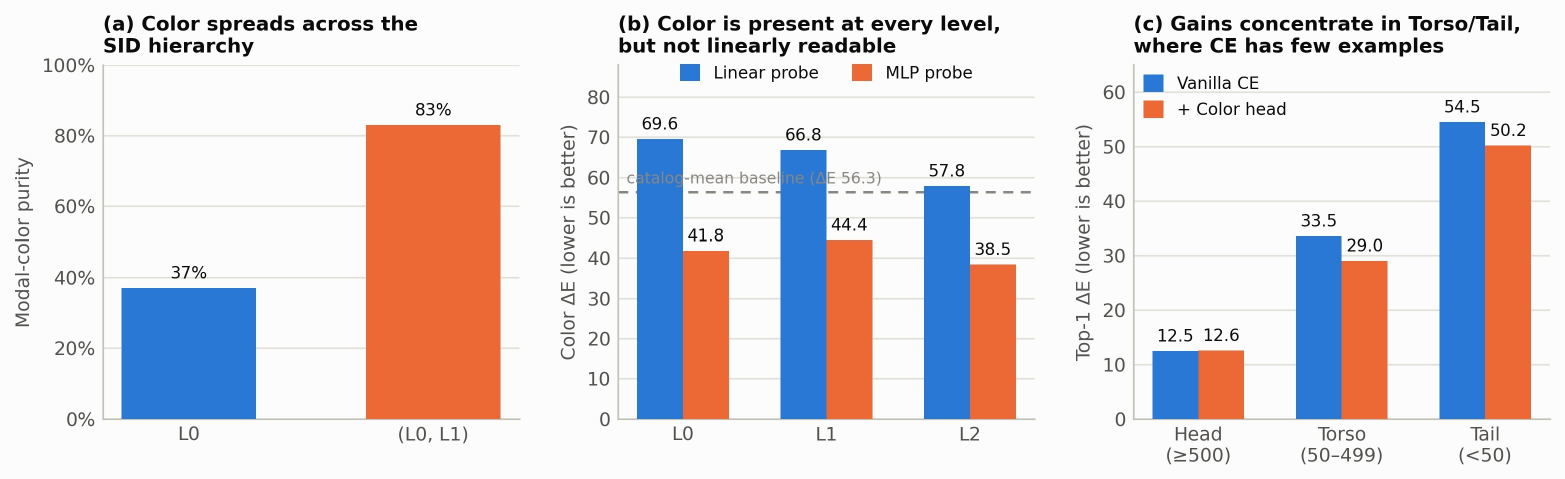}
  \caption{Motivation for attribute-specific supervision. (a) Color purity increases across SID levels, revealing fine-grained attribute structure. (b) Nonlinear probes decode color better than linear probes, indicating latent but unorganized attribute information. (c) Attribute supervision mainly reduces color prediction error for torso and tail items with sparse interaction data.}
  \label{fig:attr_mot}
  \Description{bar charts comparing std CE \& hierarchical loss.}
  
\end{figure}

Color provides a concrete example of this limitation. In our catalog, each item is associated with a CIE-LAB color representation, allowing us to quantify perceptual similarity between items. As shown in Figure~\ref{fig:attr_mot}, color information is distributed across the SID hierarchy: the L0 code alone achieves only $37\%$ mean color purity, whereas conditioning on the $(L0,L1)$ prefix increases color purity to $83\%$. This indicates that fine-grained color information emerges progressively across the hierarchy rather than being localized to a single SID level. Moreover, a probe analysis shows that color information is present in the model's hidden representations but is not readily accessible through a simple linear mapping. While the linear probe performs worse than the catalog-mean baseline at all SID positions, a lightweight MLP substantially improves color prediction, achieving a $\Delta E$ of $38.5$ at L2 compared with the baseline of $56.3$. This suggests that attribute information exists in the representation but is not explicitly organized for direct attribute prediction.

The limitation is particularly important for long-tail items, which receive substantially less interaction supervision. In our catalog, $15.1\%$ of items occur at most five times in the training data, while $23.8\%$ never appear in a color-specific conversation, leaving them with little or no direct color supervision. 


As a result, a model may learn to generate plausible semantic IDs while failing to preserve fine-grained attributes needed to distinguish similar items.

These observations motivate an additional source of \textbf{attribute-specific supervision}. Rather than relying solely on semantic-ID prediction to implicitly encode every item property, we explicitly encourage the model's hidden representations to preserve fine-grained attribute information. Using color as a concrete instantiation, we apply auxiliary supervision to the item's catalog-derived CIE-LAB representation. As shown in Figure~\ref{fig:attr_mot}, this supervision primarily benefits data-sparse items: the mean top-1 color error decreases from $54.5$ to $50.2$ for tail colors and from $33.5$ to $29.0$ for torso colors, while head colors show essentially no improvement ($12.5$ to $12.6$). This suggests that attribute supervision provides complementary information precisely where interaction-based supervision is weakest. Importantly, the formulation is not specific to color and can be instantiated with other item attributes, such as material, fit, or pattern.

\section{Methodology}
\label{sec:method_overview}

We propose \textbf{HASTE}, a unified fine-tuning framework that improves semantic-ID-based generative recommendation through two complementary forms of supervision. Given a user interaction context, the model autoregressively generates an item's semantic ID as a sequence of discrete codes obtained through residual quantization. While standard causal cross-entropy provides supervision for predicting these codes, it treats the semantic-ID sequence uniformly and does not explicitly account for its hierarchical structure. Moreover, item-level attributes provide complementary semantic information that can be particularly valuable for learning representations of data-sparse tail items. Figure~\ref{fig:overview} provides an overview of the proposed training framework.


HASTE introduces two auxiliary objectives to address these limitations. The first is a \textbf{Hierarchical Loss}, which explicitly exploits the multi-level structure of semantic IDs and provides level-aware supervision during semantic-ID generation. By aligning the training objective with the hierarchical organization of the semantic-ID space, the model can learn coarse-to-fine item semantics more effectively. The second is an \textbf{Attribute-Specific Loss}, which introduces auxiliary supervision from fine-grained item attributes. The loss encourages intermediate model representations to retain attribute-level information that may not be sufficiently captured by the semantic-ID prediction objective alone. This additional supervision is particularly beneficial for tail items, where limited interaction data makes it difficult to learn informative representations solely from the primary recommendation objective.

The overall HASTE objective combines the standard causal language-model loss with these two complementary sources of supervision:
\begin{equation}
\mathcal{L}_{\mathrm{HASTE}}
=
\mathcal{L}_{\mathrm{hier}}
+
\lambda_{\mathrm{attr}}\mathcal{L}_{\mathrm{attr}},
\end{equation}
where $\lambda_{\mathrm{attr}}$ controls the contribution of the attribute-specific objective. The hierarchical loss serves as the primary semantic-ID prediction objective, replacing the standard causal cross-entropy loss, while the attribute loss provides complementary supervision for learning fine-grained item representations. Both the backbone and the auxiliary attribute prediction head are jointly optimized.

\subsection{Hierarchical SID Loss}
\label{sec:hierarchical_loss}

Let an item be represented by a four-level Semantic ID,
$\mathbf{y}=(y_1,y_2,y_3,\\y_4)$, where $y_\ell \in V_\ell$ denotes the
code at hierarchy level $\ell$, ordered from coarse to fine. The model
autoregressively predicts each SID token, with $y_2$ conditioned on
$y_1$, with $y_3$ conditioned on $(y_1,y_2)$ and with $y_4$ conditioned on $(y_1,y_2,y_3)$. Unlike standard
token-level cross-entropy, which treats all SID positions equally, we
introduce a hierarchy-aware loss that accounts for both the structural
importance of each level and the model's confidence in its upstream
predictions.

\noindent\paragraph{Hierarchical weighting.}
We assign decreasing base weights to successive SID levels using a simple linear schedule,
\begin{equation}
(w_1,w_2,w_3,w_4)=(4,3,2,1),
\end{equation}
which provides a moderate bias toward coarser levels while retaining supervision at finer levels. Errors at coarser levels can redirect generation to an incorrect subtree and therefore affect downstream predictions; accordingly, they receive larger weights. The root level $L_1$ is not gated since it has no upstream SID token.

\noindent \paragraph{Confidence-aware conditional weighting.}
For each SID position $t$, we compute the model's confidence in the
ground-truth token as
\begin{equation}
p_t =
\left[
\max\left(
\operatorname{softmax}(\mathbf{z}_t)_{y_t},
\epsilon
\right)
\right]^{\gamma},
\end{equation}
where $\mathbf{z}_t$ denotes the model logits, $\epsilon$ is a minimum
confidence floor, and $\gamma$ controls the sharpness of the confidence
signal. The confidence is detached from the computation graph to prevent
the model from manipulating its loss weight.

The confidence gates for the four hierarchy levels are defined as
\begin{equation}
g_1 = 1,\qquad
g_2 = p_1,\qquad
g_3 = p_1p_2, \qquad
g_4 = p_1p_2p_3.
\end{equation}
Thus, supervision at a downstream level is strengthened when the model
is confident in the preceding decisions, while the confidence floor
ensures that a minimum learning signal is retained when upstream
predictions are uncertain.

\noindent \paragraph{Annealed hierarchical-to-conditional weighting.}
To avoid relying on unreliable confidence estimates at the beginning
of training, we gradually transition from static hierarchical weighting
to confidence-aware weighting. Let $\alpha$ be linearly annealed from
$1$ to $0$. The effective weight for level $\ell$ is
\begin{equation}
\tilde{w}_{\ell}
=
w_{\ell}
\left[
\alpha + (1-\alpha)g_{\ell}
\right].
\end{equation}
At the beginning of training ($\alpha=1$), the loss uses only the
stable hierarchical weights. As training progresses, the contribution
of the confidence gate increases, reaching fully conditional weighting
when $\alpha=0$.

\noindent \paragraph{Scheduled sampling.}
Autoregressive SID generation introduces a train--inference mismatch:
during teacher forcing, downstream levels are conditioned on
ground-truth upstream tokens, whereas inference conditions on the
model's own predictions. We reduce this mismatch through scheduled
sampling. With probability $p_{\mathrm{ss}}$, the ground-truth $L_1$
token is replaced by a token sampled from the model's predicted
distribution. The sampling probability is linearly increased from zero
to a maximum value over the scheduled-sampling annealing period.

The sampled prefix is then used by the subsequent loss-computing
forward pass. Consequently, the model is exposed to the types of
upstream prediction errors encountered during autoregressive generation,
rather than being trained exclusively on clean, teacher-forced
trajectories.

\noindent \paragraph{Overall objective.}
The resulting hierarchical SID loss is
\begin{equation}
\mathcal{L}_{\mathrm{H}}
=
\frac{
\sum_t
\tilde{w}_t\,
\mathcal{L}_{\mathrm{CE}}(t)\,
\mathbf{1}[\operatorname{valid}(t)]
}{
\sum_t \mathbf{1}[\operatorname{valid}(t)]
},
\end{equation}
where $\tilde{w}_t$ is the effective weight associated with the SID
level of position $t$, and $\mathbf{1}[\operatorname{valid}(t)]$
masks non-label positions.

Together, these components provide structure- and
confidence-aware supervision: coarse SID decisions receive greater
emphasis, while fine-grained supervision is progressively conditioned
on the model's ability to resolve upstream semantic decisions. Since
both confidence estimation and the final loss are computed on the same
(possibly perturbed) input trajectory, the resulting weighting remains
consistent with the conditioning context seen by the model.

\subsection{Attribute-Conditioned Auxiliary Losses}
\label{sec:attribute_loss}

Hierarchical supervision encourages the model to preserve the coarse-to-fine structure of Semantic IDs (SIDs), but it does not explicitly constrain the learned representations to retain fine-grained item attributes. This is particularly important for attributes such as color and material composition, which may be weakly supervised by conversational interactions, especially for infrequently occurring items. To address this limitation, we introduce a generic attribute-aware auxiliary objective that encourages the hidden representation used for fine-grained SID prediction to preserve attribute information. The proposed objective is not specific to any particular attribute and can be instantiated for any attribute that can be represented by a suitable prediction target. We demonstrate its effectiveness using two complementary item attributes, color and material composition.

\begin{table*}[h]
\centering
\caption{Performance comparison across different recommendation methods. 
The best result is shown in \textbf{bold} and the second-best result is 
\underline{underlined}.}
\label{tab:main_results}
\small
\setlength{\tabcolsep}{2pt}
\renewcommand{\arraystretch}{1.15}

\begin{tabular}{llcccccccc}
\toprule
\textbf{Dataset} & \textbf{Metric} &
\multicolumn{2}{c}{\textbf{Sequential}} &
\multicolumn{2}{c}{\textbf{Generative}} &
\multicolumn{2}{c}{\textbf{LLM-Based}} &
\textbf{Ours (CE loss)} & \textbf{Ours (Hier loss)} \\
\cmidrule(lr){3-4}
\cmidrule(lr){5-6}
\cmidrule(lr){7-8}
&
& \textbf{SASRec~\cite{kang2018selfattentivesequentialrecommendation}} & \textbf{BERT4Rec~\cite{sun2019bert4recsequentialrecommendationbidirectional}}
& \textbf{TIGER~\cite{rajput2023recommendersystemsgenerativeretrieval}} & \textbf{HSTU~\cite{zhai2024actionsspeaklouderwords}}
& \textbf{IDGenRec~\cite{tan2024idgenrecllmrecsysalignmenttextual}} & \textbf{OneRec-Think~\cite{liu2025onerecthinkintextreasoninggenerative}}
& \textbf{} & \textbf{} \\
\midrule

\multirow{4}{*}{\textbf{Beauty}}
& Recall@5
& 0.0402 & 0.0232 & 0.0405 & 0.0424
& 0.0484 & \underline{0.0563} & 0.0325 & \textbf{0.0588} {\scriptsize\textcolor{blue}{(+4.4\%)}} \\
& Recall@10
& 0.0607 & 0.0396 & 0.0623 & 0.0652
& 0.0693 & \underline{0.0791} & 0.0672 & \textbf{0.0823} {\scriptsize\textcolor{blue}{(+4.0\%)}} \\
& NDCG@5
& 0.0254 & 0.0146 & 0.0267 & 0.0280
& 0.0337 & \textbf{0.0398} & 0.0326 & \underline{0.0387} {\scriptsize\textcolor{blue}{(-2.8\%)}} \\
& NDCG@10
& 0.0320 & 0.0199 & 0.0337 & 0.0353
& 0.0404 & \underline{0.0471} & 0.0358 & \textbf{0.0503} {\scriptsize\textcolor{blue}{(+6.8\%)}}\\

\midrule

\multirow{4}{*}{\textbf{Sports}}
& Recall@5
& 0.0199 & 0.0102 & 0.0215 & 0.0268
& 0.0270 & \underline{0.0288} & 0.0249 &\textbf{0.0322} {\scriptsize\textcolor{blue}{(+11.8\%)}}\\
& Recall@10
& 0.0301 & 0.0175 & 0.0347 & 0.0343
& 0.0388 & \underline{0.0412} & 0.0327 & \textbf{0.0463} {\scriptsize\textcolor{blue}{(+12.4\%)}} \\
& NDCG@5
& 0.0106 & 0.0065 & 0.0137 & 0.0173
& 0.0185 & \textbf{0.0199} & 0.0131 & \underline{0.0188} {\scriptsize\textcolor{blue}{(-5.5\%)}}\\
& NDCG@10
& 0.0141 & 0.0088 & 0.0179 & 0.0226
& 0.0223 & \underline{0.0239} & 0.0212 & \textbf{0.0279} {\scriptsize\textcolor{blue}{(+16.7\%)}}\\

\midrule

\multirow{4}{*}{\textbf{Toys}}
& Recall@5
& 0.0448 & 0.0215 & 0.0337 & 0.0366
& \underline{0.0595} & 0.0579 & 0.0396 & \textbf{0.0655} {\scriptsize\textcolor{blue}{(+10.1\%)}}\\
& Recall@10
& 0.0626 & 0.0332 & 0.0547 & 0.0566
& \underline{0.0800} & 0.0797 & 0.0571 & \textbf{0.0924} {\scriptsize\textcolor{blue}{(+15.5\%)}} \\
& NDCG@5
& 0.0300 & 0.0131 & 0.0209 & 0.0245
& \underline{0.0432} & 0.0412 & 0.0272 & \textbf{0.0479} {\scriptsize\textcolor{blue}{(+10.9\%)}}\\
& NDCG@10
& 0.0358 & 0.0168 & 0.0276 & 0.0309
& \textbf{0.0498} & 0.0482 & 0.0328 & \underline{0.0492} {\scriptsize\textcolor{blue}{(-1.2\%)}} \\

\bottomrule
\end{tabular}
\end{table*}

\noindent \paragraph{Auxiliary prediction heads.}
For each attribute $a$, we attach a lightweight MLP regression head
[
$f\_a:\mathbb{R}^{d}\rightarrow\mathbb{R}^{d\_a}$
],
with architecture $\text{Linear}(d,64)$--GELU--$\text{Linear}(64,d_a)$, where $d$ is the backbone hidden dimension and $d_a$ is the dimensionality of the attribute target. The head operates on the final-layer hidden state at the position used to predict the third-level SID token $L_2$. At this position, the model has already conditioned on the preceding coarse-grained codes $L_0$ and $L_1$, making it a natural location for capturing fine-grained perceptual and compositional properties of the item.

\noindent \paragraph{SID-conditioned attribute targets.}
Attribute targets are constructed independently of the SID token vocabulary. We first join the production catalog, which associates each item with its complete SID tuple $(l_0,l_1,l_2,l_3)$, with the corresponding raw attribute annotations. Since multiple items may share the same SID tuple, we aggregate the attribute annotations of all items assigned to a tuple to obtain a lookup function
$[
T_a(l_0,l_1,l_2,l_3)\rightarrow\mathbf{y}_a
]$.
During training, the attribute target is therefore determined directly from the item's complete SID, rather than from the conversation or the surface form of the user query. This allows the auxiliary objective to be applied to every valid SID-generation example, including examples whose conversational context does not explicitly mention the attribute.

\noindent \paragraph{Position-aligned supervision.}
The attribute head operates on the final-layer hidden state at the $L_2$ SID prediction position, which is the same representation used by the backbone to predict the fine-grained SID token after conditioning on the preceding $L_0$ and $L_1$ codes. We choose this position because the finer-grained information captured at the $L_2$ level is expected to be more closely related to perceptual and compositional item attributes. As discussed in Section~\ref{sec:motivation}, our preliminary probing experiments provide empirical evidence that the $L_2$ representation contains substantially more attribute information than earlier SID positions. Thus, aligning the auxiliary supervision with the $L_2$ prediction position directly encourages the representation used for fine-grained item prediction to preserve attribute-specific information.
\[
\mathcal{L}_{\mathrm{HASTE}} = \mathcal{L}_{\mathrm{gen}} + \lambda_a \mathcal{L}_a,
\]
where $\mathcal{L}_{\mathrm{gen}}$ denotes the generative SID objective and $\lambda_a$ controls the strength of attribute supervision.

The auxiliary heads are applied to all eligible SID-generation examples rather than only attribute-specific conversations. Because the target is obtained through the SID-to-attribute lookup, the supervision is independent of task type and conversational wording. This substantially increases the density of attribute supervision and is particularly beneficial for items that occur infrequently in attribute-specific interactions.

\noindent \paragraph{Attribute-specific instantiations.}
We instantiate the framework for two complementary item attributes: color and material composition.

\textbf{Color.} For color, the auxiliary head predicts the item's CIELAB~\cite{CIE1976Lab} color representation, normalized as

\[
\left[
\frac{L^*}{50},
\frac{a^*}{80},
\frac{b^*}{80}
\right].
\]

and use a Huber loss with $\delta=0.1$. A preliminary frozen-backbone probe study identified the $L_2$ prediction position as the most informative attachment point, substantially reducing color error compared with earlier SID positions. The same study showed that a nonlinear MLP probe outperforms a linear probe, suggesting that color information is encoded in the representation but is not linearly organized.

To further address the limited supervision available for infrequent items, we apply frequency-aware weighting to the color loss. For an item with training frequency $f_i$, its auxiliary loss is scaled by
\[
r_i =
\operatorname{clamp}
\left(
\left(\frac{f_{\mathrm{med}}}{f_i}\right)^\gamma,
1,5
\right),\qquad
\gamma=0.5
\]

where $f_{\mathrm{med}}$ is the median item frequency. Thus, low-frequency items receive stronger attribute supervision while preventing excessively large per-example weights.

\textbf{Material.}
For material, the target is a 15-dimensional composition vector consisting of 14 named fiber categories and an ``other'' category. We parse free-text material annotations using a rule-based normalization procedure that incorporates explicit percentages and distributes remaining mass across unweighted material mentions, followed by $\ell_1$ normalization:
$[
\sum_{k=1}^{15} y_{\mathrm{mat},k}=1.
]$
The resulting targets lie in $[0,1]$, so no additional normalization is required. We use the same Huber objective with $\delta=0.1$.

\noindent \paragraph{Decoupling from generation.}
The auxiliary objectives do not modify the SID vocabulary, logits, or token-prediction distribution. Instead, they impose additional supervision on the shared hidden representation through lightweight task-specific heads. Consequently, the generative objective remains unchanged while the backbone is encouraged to encode attribute structure that may otherwise be weakly supervised by interaction data. The auxiliary heads are maintained separately from the backbone checkpoint and saved with their architecture metadata and parameters.

\begin{table*}[h]
\centering
\caption{Overall and tail-item recommendation performance under hierarchical and attribute-specific supervision. Values in blue denote relative improvement over the corresponding baseline: CE for HL and CE-based attribute variants, and HL for HL-based attribute variants. Bold values indicate the best performance, while \underline{underlined} values indicate the second-best performance.}
\label{tab:tail_ablation}
\resizebox{\linewidth}{!}{
\begin{tabular}{llcccccccc}
\toprule
\textbf{Method} & \textbf{Split}
& \multicolumn{4}{c}{\textbf{Amazon Fashion-140k}}
& \multicolumn{4}{c}{\textbf{Internal-fashion}} \\
\cmidrule(lr){3-6}
\cmidrule(lr){7-10}
& & 
\multicolumn{2}{c}{\textbf{Recall}}
& \multicolumn{2}{c}{\textbf{NDCG}}
& \multicolumn{2}{c}{\textbf{Recall}}
& \multicolumn{2}{c}{\textbf{NDCG}} \\
& &
\textbf{@5} & \textbf{@10}
& \textbf{@5} & \textbf{@10}
& \textbf{@5} & \textbf{@10}
& \textbf{@5} & \textbf{@10} \\
\midrule

CE
& Overall
& 0.0243 & 0.0324 & 0.0236 & 0.0452
& 0.0255 & 0.0319 & 0.0376 & 0.0425 \\
& Tail
& 0.0134 & 0.0169 & 0.0245 & 0.0298
& 0.0267 & 0.0342 & 0.0387 & 0.0393 \\

\addlinespace

HL
& Overall
& 0.0297 {\scriptsize\textcolor{blue}{(+22.2\%)}}
& 0.0345 {\scriptsize\textcolor{blue}{(+6.5\%)}}
& 0.0316 {\scriptsize\textcolor{blue}{(+33.9\%)}}
& 0.0486 {\scriptsize\textcolor{blue}{(+7.5\%)}}
& 0.0284 {\scriptsize\textcolor{blue}{(+11.4\%)}}
& 0.0351 {\scriptsize\textcolor{blue}{(+10.0\%)}}
& 0.0412 {\scriptsize\textcolor{blue}{(+9.6\%)}}
& 0.0468 {\scriptsize\textcolor{blue}{(+10.1\%)}} \\

& Tail
& 0.0168 {\scriptsize\textcolor{blue}{(+25.4\%)}}
& 0.0215 {\scriptsize\textcolor{blue}{(+27.2\%)}}
& 0.0297 {\scriptsize\textcolor{blue}{(+21.2\%)}}
& 0.0354 {\scriptsize\textcolor{blue}{(+18.8\%)}}
& 0.0301 {\scriptsize\textcolor{blue}{(+12.7\%)}}
& 0.0387 {\scriptsize\textcolor{blue}{(+13.2\%)}}
& 0.0435 {\scriptsize\textcolor{blue}{(+12.4\%)}}
& 0.0459 {\scriptsize\textcolor{blue}{(+16.8\%)}} \\

\addlinespace

CE + Color Loss
& Overall
& 0.0256 {\scriptsize\textcolor{blue}{(+5.3\%)}}
& 0.0335 {\scriptsize\textcolor{blue}{(+3.4\%)}}
& 0.0251 {\scriptsize\textcolor{blue}{(+6.4\%)}}
& 0.0464 {\scriptsize\textcolor{blue}{(+2.7\%)}}
& 0.0270 {\scriptsize\textcolor{blue}{(+5.9\%)}}
& 0.0331 {\scriptsize\textcolor{blue}{(+3.8\%)}}
& 0.0391 {\scriptsize\textcolor{blue}{(+4.0\%)}}
& 0.0438 {\scriptsize\textcolor{blue}{(+3.1\%)}} \\

& Tail
& 0.0167 {\scriptsize\textcolor{blue}{(+24.6\%)}}
& 0.0218 {\scriptsize\textcolor{blue}{(+29.0\%)}}
& 0.0298 {\scriptsize\textcolor{blue}{(+21.6\%)}}
& 0.0361 {\scriptsize\textcolor{blue}{(+21.1\%)}}
& 0.0319 {\scriptsize\textcolor{blue}{(+19.5\%)}}
& 0.0412 {\scriptsize\textcolor{blue}{(+20.5\%)}}
& 0.0461 {\scriptsize\textcolor{blue}{(+19.1\%)}}
& 0.0487 {\scriptsize\textcolor{blue}{(+23.9\%)}} \\

\addlinespace

HL + Color Loss
& Overall
& \textbf{0.0305} {\scriptsize\textcolor{blue}{(+2.7\%)}}
& \textbf{0.0353} {\scriptsize\textcolor{blue}{(+2.3\%)}}
& \textbf{0.0325} {\scriptsize\textcolor{blue}{(+2.8\%)}}
& \textbf{0.0495} {\scriptsize\textcolor{blue}{(+1.9\%)}}
& \textbf{0.0291} {\scriptsize\textcolor{blue}{(+2.5\%)}}
& \textbf{0.0359} {\scriptsize\textcolor{blue}{(+2.3\%)}}
& \textbf{0.0423} {\scriptsize\textcolor{blue}{(+2.7\%)}}
& \textbf{0.0476} {\scriptsize\textcolor{blue}{(+1.7\%)}} \\

& Tail
& \textbf{0.0202} {\scriptsize\textcolor{blue}{(+20.2\%)}}
& \textbf{0.0267} {\scriptsize\textcolor{blue}{(+24.2\%)}}
& \textbf{0.0358} {\scriptsize\textcolor{blue}{(+20.5\%)}}
& \textbf{0.0431} {\scriptsize\textcolor{blue}{(+21.8\%)}}
& \textbf{0.0348} {\scriptsize\textcolor{blue}{(+15.6\%)}}
& \textbf{0.0456} {\scriptsize\textcolor{blue}{(+17.8\%)}}
& \textbf{0.0517} {\scriptsize\textcolor{blue}{(+18.9\%)}}
& \textbf{0.0549} {\scriptsize\textcolor{blue}{(+19.6\%)}} \\

\addlinespace

CE + Material Loss
& Overall
& 0.0252 {\scriptsize\textcolor{blue}{(+3.7\%)}}
& 0.0332 {\scriptsize\textcolor{blue}{(+2.5\%)}}
& 0.0248 {\scriptsize\textcolor{blue}{(+5.1\%)}}
& 0.0461 {\scriptsize\textcolor{blue}{(+2.0\%)}}
& 0.0268 {\scriptsize\textcolor{blue}{(+5.1\%)}}
& 0.0328 {\scriptsize\textcolor{blue}{(+2.8\%)}}
& 0.0389 {\scriptsize\textcolor{blue}{(+3.5\%)}}
& 0.0435 {\scriptsize\textcolor{blue}{(+2.4\%)}} \\

& Tail
& 0.0161 {\scriptsize\textcolor{blue}{(+20.1\%)}}
& 0.0209 {\scriptsize\textcolor{blue}{(+23.7\%)}}
& 0.0287 {\scriptsize\textcolor{blue}{(+17.1\%)}}
& 0.0348 {\scriptsize\textcolor{blue}{(+16.8\%)}}
& 0.0310 {\scriptsize\textcolor{blue}{(+16.1\%)}}
& 0.0401 {\scriptsize\textcolor{blue}{(+17.3\%)}}
& 0.0450 {\scriptsize\textcolor{blue}{(+16.3\%)}}
& 0.0478 {\scriptsize\textcolor{blue}{(+21.6\%)}} \\

\addlinespace

HL + Material Loss
& Overall
& \underline{0.0303} {\scriptsize\textcolor{blue}{(+2.0\%)}}
& \underline{0.0351} {\scriptsize\textcolor{blue}{(+1.7\%)}}
& \underline{0.0322} {\scriptsize\textcolor{blue}{(+1.9\%)}}
& \underline{0.0492} {\scriptsize\textcolor{blue}{(+1.2\%)}}
& \underline{0.0289} {\scriptsize\textcolor{blue}{(+1.8\%)}}
& \underline{0.0357} {\scriptsize\textcolor{blue}{(+1.7\%)}}
& \underline{0.0420} {\scriptsize\textcolor{blue}{(+1.9\%)}}
& \underline{0.0473} {\scriptsize\textcolor{blue}{(+1.1\%)}} \\

& Tail
& \underline{0.0195} {\scriptsize\textcolor{blue}{(+16.1\%)}}
& \underline{0.0258} {\scriptsize\textcolor{blue}{(+20.0\%)}}
& \underline{0.0349} {\scriptsize\textcolor{blue}{(+17.5\%)}}
& \underline{0.0419} {\scriptsize\textcolor{blue}{(+18.4\%)}}
& \underline{0.0341} {\scriptsize\textcolor{blue}{(+13.3\%)}}
& \underline{0.0447} {\scriptsize\textcolor{blue}{(+15.5\%)}}
& \underline{0.0508} {\scriptsize\textcolor{blue}{(+16.8\%)}}
& \underline{0.0538} {\scriptsize\textcolor{blue}{(+17.2\%)}} \\

\bottomrule
\end{tabular}
}
\end{table*}

\section{Experiments}


\textbf{Implementation Details.}
\label{sec:imple_details} 
Each semantic ID consists of four tokens $(L_0,L_1,L_2,L_3)$, with a 256-entry RQ-VAE codebook at each level. Our hierarchical loss uses level-specific base weights $(4.0, 3.0, \\ 2.0, 1.0)$, confidence-based gating, an annealed transition from static weighting to confidence gating, and scheduled sampling to reduce the train--inference exposure gap. For attribute supervision, we use a CIE-LAB distance-based soft-label objective at $L_2$, restricted to candidates sharing the ground-truth $(L_0,L_1)$ prefix, and apply it to color-conditioned conversations with weight $0.1$. We further employ a hidden-state regression head to provide complementary attribute supervision. Training is performed in two stages, with embedding warm-up followed by full fine-tuning using learning rate $2\times10^{-5}$, effective batch size 32, and a cosine learning-rate schedule. We train the Qwen3-8B model in two stages. Stage 1 extends the model vocabulary with semantic-ID (SID) tokens and initializes their embeddings. Stage 2 performs full-parameter supervised fine-tuning (SFT), initialized from the Stage 1 checkpoint. All experiments were conducted on NVIDIA RTX6000 single GPU instance with 96 GB memory. Additional implementation details and hyperparameters are provided in the Supplementary Material.

\noindent \textbf{Datasets.}
\label{sec:datasets}
We evaluate HASTE across diverse recommendation domains and data regimes. We use four public domains from the Amazon Reviews benchmark~\cite{hou2024bridging} -- Sports \& Outdoors (Sports), Beauty, Toys \& Games (Toys), and Fashion-140k, a 140K-item subset of the larger Amazon Fashion dataset, and an internal production-scale proprietary fashion dataset (internal-fashion). Sports, Beauty, and Toys follow the standard P5 5-core protocol \cite{geng2022p5}, where each user and item has at least five interactions, and are used to evaluate the effectiveness of hierarchical supervision for sequential generative recommendation. Fashion-140k and internal-fashion dataset provide richer item-level attribute annotations, including color and material, enabling evaluation of the proposed attribute-specific supervision, particularly in the long-tail setting. Across datasets, item representations are encoded as hierarchical semantic IDs using a 3-level RQ-VAE with 256 codes per level, with a deduplication token appended where applicable. Further details on dataset statistics, preprocessing, semantic-ID construction, and attribute availability are provided in the supplementary material.

\noindent \textbf{Baselines.} We compare HASTE against representative baselines spanning three categories: traditional sequential recommendation methods (BERT4Rec~\cite{sun2019bert4recsequentialrecommendationbidirectional}, SASRec~\cite{kang2018selfattentivesequentialrecommendation}), generative recommendation methods (TIGER~\cite{rajput2023recommendersystemsgenerativeretrieval}, HSTU~\cite{zhai2024actionsspeaklouderwords}), and LLM-based methods (IDGenRec~\cite{tan2024idgenrecllmrecsysalignmenttextual}, OneRec-Think~\cite{liu2025onerecthinkintextreasoninggenerative}). We additionally compare against the standard cross-entropy (CE) objective to isolate the impact of our proposed supervision.

\noindent \textbf{Evaluation Protocol.} We use $\mathrm{Recall@K}$ and $\mathrm{NDCG@K}$ with $K \in \{5,10\}$ to evaluate recommendation performance. We evaluate the hierarchical loss on the sequential recommendation task, while attribute-specific losses are evaluated on attribute-conditioned product recommendation, both overall and on torso and tail items with fewer than 500 instances.

\subsection{Results}
We compare HASTE against representative sequential, generative, and LLM-based recommendation baselines, and evaluate the contribution of its hierarchical and attribute-specific supervision through complimentary experiments. 

\noindent \textbf{Overall recommendation performance across Beauty, Sports, and Toys.} Table~\ref{tab:main_results} compares HASTE with baseline methods across Beauty, Sports, and Toys datasets. HASTE with hierarchical supervision consistently achieves strong performance, particularly on Recall@10, improving over the strongest baseline by 4.0\%. 12.4\%, and 15.5\% on Beauty, Sports, and Toys, respectively. Similar gains are observed for NDCG@10, with improvements of 6.8\%, 16.7\%, and 1.2\%. The stronger gains on Sports and Toys suggest that hierarchical supervision effectively captures the structured relationships encoded by semantic IDs. Compared with the CE-based variant, hierarchical supervision generally provides further improvements, demonstrating the benefit of explicitly supervising intermediate semantic-ID levels.

\noindent \textbf{Tail-item and attribute-aware performance.} Table~\ref{tab:tail_ablation} evaluates hierarchical and attribute-specific supervision on Amazon Fashion-140k and the production-scale proprietary internal-fashion dataset. We use these datasets for attribute losses because Toys, Sports, and Beauty lack the required item-level attribute annotations. Hierarchical supervision consistently improves over CE, with larger gains on tail items, while color and material supervision provide further improvements, particularly for tail items. Under HL, color supervision gives the strongest gains, reaching 24.2\% in Recall@10 on Fashion-140k and 17.8\% on internal-fashion, while achieving the best overall and tail performance on internal-fashion. These results show that hierarchical and attribute-specific supervision are complementary: the former improves semantic retrieval, while the latter preserves fine-grained attribute information valuable for tail items.

\begin{table}[h]
\centering
\caption{Training efficiency of standard causal cross-entropy (CE) and Hierarchical Loss (HL). HL substantially reduces the number of training steps required to reach convergence.}
\label{tab:convergence}
\begin{tabular}{lccc}
\toprule
\textbf{Dataset} & \textbf{CE Steps} & \textbf{HL Steps} &
\textbf{Speedup} \\
\midrule
Toys & 92000 & 42000 & 2.19$~\times$ \\
Sports & 100000 & 53000 & 1.89$~\times$ \\
Beauty & 85000 & 44000 & 1.93$~\times$ \\
\midrule
\textbf{Average} & -- & -- & \textbf{$1.99~\times$} \\
\bottomrule
\end{tabular}
\end{table}

\noindent \textbf{Training efficiency.} Table 3 compares the optimization speed of standard causal CE and HL across three Amazon datasets. We define convergence as reaching the same validation-loss level achieved by the corresponding CE model, rather than directly comparing the raw CE and HL loss values. HL consistently reaches this target in substantially fewer training steps, reducing the required steps by more than half on Toys and by roughly half on Sports and Beauty. On average, HL achieves a 1.99$\times$ speedup over CE, demonstrating that hierarchical supervision not only improves recommendation quality but also enables more efficient optimization.

\noindent \textbf{Ablation Studies.} 
We conduct ablation on the Toys dataset to examine the contribution of hierarchical supervision and its individual components. Table~\ref{tab:sid_level_ablation} shows that progressively supervising additional semantic-ID levels consistently improves NDCG@10. Performance increases from 0.0355 with supervision at the first level to 0.0492 when all four levels are supervised, demonstrating that finer-grained semantic codes provide complementary supervision. 

Table~\ref{tab:hier_component_ablation} further isolates the contributions of the hierarchical loss components. Static weighting (SW) assigns fixed weights of $4{:}3{:}2{:}1$ to the four SID levels, emphasizing coarser semantic decisions. Conditional gating (CG) uses confidence-aware weights, strengthening downstream supervision when preceding predictions are reliable. Scheduled sampling (SS) gradually replaces the ground-truth $L_1$ token with model predictions to reduce the teacher-forcing mismatch. Each component improves over causal CE, with CG providing the largest individual gain. The full configuration combines SW, CG, and SS, yielding the best performance and improving Recall@10 and NDCG@10 by 44.3\% and 50.0\%, respectively, over causal CE. The substantial improvement from the combined setup suggests that the three mechanisms provide complementary benefits: level-aware weighting, confidence-adaptive supervision, and robustness to autoregressive errors. Further details on the hyperparameter settings used in this ablation are provided in the supplementary material.

\begin{table}[h]
\centering
\small
\setlength{\tabcolsep}{4pt}
\caption{Ablation of hierarchical supervision across semantic-ID levels on Toys dataset. Using conditional gating, we progressively incorporate higher-level semantic codes into the Hierarchical Loss.}
\label{tab:sid_level_ablation}
\begin{tabular}{lcc}
\toprule
\textbf{SID Supervision} & \textbf{NDCG@10} & \textbf{$\Delta$ vs. CE} \\
\midrule
Causal CE & 0.0328 & -- \\
$(s_a)$ & 0.0355 & +0.0027 \\
$(s_a,s_b)$ & 0.0410 & +0.0082 \\
$(s_a,s_b,s_c)$ & 0.0435 & +0.0107 \\
$(s_a,s_b,s_c,s_d)$ & \textbf{0.0492} & \textbf{+0.0164} \\
\bottomrule
\end{tabular}
\end{table}

\begin{table}[h]
\centering
\caption{Ablation of individual components of the Hierarchical Loss on the Toys dataset.
Here, \textit{HL} denotes Hierarchical Loss, \textit{SW} static level weighting,
\textit{CG} conditional gating, and \textit{SS} scheduled sampling.}
\label{tab:hier_component_ablation}
\begin{tabular}{lccc}
\toprule
\textbf{Supervision} & \textbf{Recall@10} & \textbf{NDCG@10} & \textbf{$\Delta$ vs. CE (\%)} \\
\midrule
CE & 0.0571 & 0.0328  & -- \\
HL w/ SW & 0.0624 & 0.0386 & +9.3 / +17.7 \\
HL w/ CG & 0.0678 & 0.0402 &  +18.7 / +22.6 \\
HL w/ SS & 0.0598 & 0.0357 & +4.7 / +8.8\\
HL w/ SW, CG, SS & 0.0824 & 0.0492 & +44.3 / +50.0 \\
\bottomrule
\end{tabular}
\end{table}

\section{Conclusion}
We presented HASTE, a unified framework for generative recommendation with hierarchical and attribute-aware supervision. By exploiting the multi-level structure of semantic IDs and incorporating confidence-aware gating, level weighting, and scheduled sampling, HASTE improves hierarchical prediction. Attribute-specific auxiliary objectives further preserve fine-grained item information, particularly for tail items. Experiments across multiple Amazon domains and a large-scale fashion dataset show consistent improvements over strong baselines, with ablations confirming the complementary benefits of the proposed components.

Some limitations of our work include its dependence on the quality of the underlying semantic-ID hierarchy and the availability of reliable attribute information for attribute-specific supervision. Moreover, our evaluation primarily relies on offline metrics, which may not fully capture real-world user satisfaction and long-term engagement. Future work will explore adaptive hierarchy construction, richer multimodal and behavioral attribute supervision, and online evaluation incorporating user feedback and long-term engagement.



\section{Ethical Considerations}

HASTE is designed to improve generative recommendation and does not directly use sensitive user attributes. However, recommendation systems can amplify existing popularity biases and may disproportionately favor frequently interacted-with items, potentially reducing exposure to less popular or tail items. HASTE explicitly incorporates attribute-aware supervision to improve tail-item representation, but this does not guarantee fairness across all products or user groups. In addition, the use of catalog and behavioral data may introduce privacy concerns depending on the data collection and deployment setting. In practice, appropriate data anonymization, access controls, and privacy-preserving data practices should be applied. Finally, the framework should be evaluated for unintended recommendation biases before deployment, particularly in settings where recommendations can influence user choices or commercial outcomes.

\section{Appendix}

\subsection{Dataset Details}
\label{sec:supp_datasets}

We evaluate HASTE on five datasets spanning public Amazon recommendation~\cite{hou2024bridging} and a production-scale internal fashion dataset: Sports \& Outdoors (Sports), Beauty, Toys \& Games (Toys), Fashion-140k, and Internal-fashion. Table~\ref{tab:dataset_statistics} summarizes their catalog sizes, user sequences, interaction volumes, conversational training data, and available metadata. Detailed preprocessing, semantic-ID construction, and attribute availability are described below.

\begin{table*}[h]
\centering
\caption{Dataset statistics. We report catalog size, user sequences, interaction events, conversation statistics, and metadata attributes. ``--'' indicates that the corresponding statistic is not applicable.}
\label{tab:dataset_statistics}
\small
\setlength{\tabcolsep}{1pt}
\begin{tabular}{lrrrrr}
\toprule
\textbf{Statistic} & \textbf{Toys} & \textbf{Beauty} & \textbf{Sports} &
\textbf{Fashion-140k} & \textbf{Internal-fashion} \\
\midrule
Items (catalog)
& 11,924 & 12,101 & 18,357 & 141,300 & 163,724 \\
User sequences
& 19,412 & 22,363 & 35,598 & 21,326 & 103,047 \\
Total events
& 167,597 & 198,502 & 296,337 & 71,059 & 1,640,953 \\
Seq. length (min/med/mean/max)
& 5/6/8.63/550
& 5/6/8.88/204
& 5/6/8.32/296
& 3/3/3.33/15
& 3/6/15.92/100 \\
Total conversations
& 1,390,005 & 1,720,660 & 2,508,966 & 1,407,009 & 103,047 \\
Train / Validation / Test
& 1,278,805/69,500/19,412
& 1,583,007/86,033/22,363
& 2,308,249/125,448/35,598
& 1,294,448/70,350/21,326
& 6,255,087/339,950/203,971 \\
Conversation types
& 22 & 26 & 26 & 40 & 40 \\
Raw metadata attributes
& 9 & 9 & 9 & 21 & 20 \\
\bottomrule
\end{tabular}
\end{table*}

\subsubsection{Sports, Beauty, and Toys.}
\label{sec:supp_sports_beauty_toys}

\paragraph{Source and preprocessing.}
Sports, Beauty, and Toys use the Amazon-2014 5-core P5 release~\cite{geng2022p5}. Each category contains one row per ASIN, while user interaction sequences are the chronological P5 5-core sequences, retaining users with at least five reviews in the corresponding category. The resulting catalogs contain 18,357, 12,101, and 11,924 items for Sports, Beauty, and Toys, respectively.

Table~\ref{tab:supp_sbt_stats} provides metadata coverage. Approximately 0.5\% of Sports and Toys items have no title in the upstream metadata (\texttt{Product: N/A}); these items produce near-degenerate embeddings and account for a disproportionate share of semantic-ID collisions. Toys has depth-3 taxonomy labels for only 28.6\% of items; therefore, statistics conditioned on this taxonomy level apply only to that subset.

\begin{table}[h]
\centering
\caption{Item and metadata coverage for Sports, Beauty, and Toys.}
\small
\begin{tabular}{lccc}
\toprule
& Sports & Beauty & Toys \\
\midrule
Items & 18,357 & 12,101 & 11,924 \\
Users (sequences) & 35,598 & 22,363 & 19,412 \\
Title coverage & 99.5\% & 99.9\% & 99.5\% \\
Description coverage & 90.5\% & 98.3\% & 98.5\% \\
Brand coverage & 67.2\% & 82.8\% & 85.6\% \\
Price coverage & 80.4\% & 95.2\% & 92.9\% \\
Taxonomy depth-3 coverage & --- & --- & 28.6\% \\
\bottomrule
\end{tabular}
\label{tab:supp_sbt_stats}
\end{table}

\paragraph{Evaluation split.}
The datasets are prepared using a leave-one-out (LOO) strategy for sequential evaluation. For each user sequence, the last item is held out for testing, the item immediately preceding it is used for validation, and all earlier items are used for training. Thus, each sequence contributes a training history, a validation target, and a test target while preserving chronological order. For the conversational SFT corpus, examples are independently shuffled and split 92\%/5\%/3\% into training, validation and test set. Sequences use a leave-two-out protocol. For each user, the training example uses seq[:-2], validation predicts seq[-2] from seq[:-2], and test predicts seq[-1] from seq[:-1].

\paragraph{Metadata and attributes.}
In addition to title, description, category, and brand, the Amazon categories provide price, sales rank, and related-item information. They do not contain the dense structured attributes required by our attribute-specific objectives, such as fine-grained color, material, style, and occasion. Consequently, these datasets are used primarily to evaluate hierarchical supervision.

\subsubsection{Fashion-140k}
\label{sec:supp_fashion140k}

\paragraph{Source and preprocessing.}
Fashion-140k contains 141,300 items selected from an internally processed 799,953-item Amazon Fashion pool. The subset is enriched with Gemini-extracted structured attributes that are not available in the raw Amazon metadata. User sequences are obtained using 5-core filtering on Amazon Fashion reviews, restricted to ASINs in the item pool. Sequences shorter than three items are removed and sequences are capped at 100 items, resulting in 21,326 sequences with mean length 3.33, median length 3, and maximum length 15.

\paragraph{Evaluation split.}
Sequences use a leave-two-out protocol. For each user, the training example uses \texttt{seq[:-2]}, validation predicts \texttt{seq[-2]} from \texttt{seq[:-2]}, and test predicts \texttt{seq[-1]} from \texttt{seq[:-1]}. All 21,326 users are retained across the splits.

\paragraph{Structured attributes.}
Fashion-140k provides dense structured attributes, including product type, gender, brand, occasion, primary and secondary color, style, material, and fit. Their coverage is summarized in Table~\ref{tab:supp_140k_stats}.

\begin{table}[h]
\centering
\caption{Structured-attribute coverage in Fashion-140k.}
\small
\begin{tabular}{lcc}
\toprule
\textbf{Attribute} & \textbf{Count} & \textbf{Coverage} \\
\midrule
Items & 141,300 & --- \\
Sequences & 21,326 & --- \\
\midrule
\texttt{product\_type} & 141,220 & 99.9\% \\
\texttt{gender} & 138,332 & 97.9\% \\
\texttt{brand} & 130,177 & 92.1\% \\
\texttt{occasion} & 130,066 & 92.0\% \\
\texttt{primary\_color} & 129,772 & 91.8\% \\
\texttt{style} & 125,727 & 89.0\% \\
\texttt{material} & 90,224 & 63.9\% \\
\texttt{secondary\_color} & 64,972 & 46.0\% \\
\texttt{fit} & 36,212 & 25.6\% \\
\bottomrule
\end{tabular}
\label{tab:supp_140k_stats}
\end{table}

\subsubsection{Internal-fashion}
\label{sec:supp_internal_fashion}

\paragraph{Source and preprocessing.}
Internal-fashion is a production-scale proprietary fashion engagement dataset containing 163,724 products, 103,047 qualifying user sequences, and 1,640,953 interaction events. The dataset is constructed from raw interaction logs joined with persona/location metadata, user-profile information, and the product catalog. Six deliberate-intent engagement types are retained: \texttt{BUY}, \texttt{LIKE}, \texttt{SHARE}, \texttt{DOWNLOAD}, \texttt{TRY\_ON}, and \texttt{SAVE}. \texttt{TAP} events are excluded because they represent a UI prerequisite rather than explicit engagement; including them would add approximately 16.5M events and cause 17\% of users to exceed the 100-item sequence cap.

Users must have a resolved city or country in the persona metadata. User identity is represented by the pair (\texttt{user\_id}, \texttt{profile\_id}), and users mapping to multiple profiles are removed. Both fully specified and partially specified persona contexts are retained during training.

\paragraph{Deduplication and sequence construction.}
Three deduplication passes remove system-generated duplicate events while preserving genuine repeated actions: ambiguous sessions containing multiple user-profile combinations are removed, exact row duplicates are discarded, and interaction-level duplicates are removed only when they match simultaneously under both session/position and timestamp-based windowing criteria. Interactions are then grouped by user-profile pair and ordered chronologically. Sequences have a minimum length of three and are capped at 100 items. Consecutive events separated by 30 days or more are discarded.

\paragraph{Training-pair generation.}
Six interaction-pattern families generate conversational training examples: cross-product transitions, same-product persistence, forward engagement, persona-conditioned prediction, backward engagement, and cross-signal escalation. Temporal gaps are bucketed into same-session ($<$30 min), within-day ($<$24h), within-week ($<$7d), and within-month (7--30d) intervals. This process produces 6,799,008 conversations, split into 6,255,087 training, 339,950 validation, and 203,971 test examples.

\paragraph{Attributes.}
The Internal-fashion catalog provides brand, price, currency, gender, color, size, materials, category name, taxonomy, and VLM/attribute-assembled product descriptions. This attribute richness makes Internal-fashion, together with Fashion-140k, the primary setting for evaluating attribute-specific supervision.

\subsubsection{Semantic-ID Construction.}
\label{sec:supp_semantic_ids}

We construct semantic IDs using a TIGER-style~\cite{rajput2023recommendersystemsgenerativeretrieval} residual vector quantization autoencoder (RQ-VAE). For Sports, Beauty, and Toys, title, description, category path, and brand are concatenated into a single context string and encoded into a 1024-dimensional representation. Fashion-140k uses the RQ-VAE checkpoint trained on the full 799,953-item Amazon Fashion pool; semantic IDs for its 141,300 items are obtained by nearest-codebook assignment in the same embedding space. Internal-fashion additionally incorporates multimodal product representations using 1024-dimensional Qwen3-VL embeddings.

The RQ-VAE encoder follows a
$1024\rightarrow512\rightarrow256\rightarrow128\rightarrow32$
architecture and uses three residual codebooks with 256 entries each. For Sports, Beauty, and Toys, the RQ-VAE is trained for 20,000 steps per category on a single A100 GPU. A fourth deduplication token is appended to the three quantized codes when collisions occur, producing the final depth-4 semantic ID used for language-model fine-tuning.

\begin{table*}[h]
\centering
\caption{Semantic-ID construction and representation statistics across datasets. Each dataset uses three residual quantization levels with 256 codes per level; a deduplication token is appended when required.}
\label{tab:dataset_sid_statistics}
\small
\setlength{\tabcolsep}{4pt}
\begin{tabular}{lccccc}
\toprule
\textbf{Statistic} & \textbf{Toys} & \textbf{Beauty} & \textbf{Sports} &
\textbf{Fashion-140k} & \textbf{Internal-fashion} \\
\midrule
Representation
& \multicolumn{3}{c}{1024-d text}
& 1024-d text
& 1024-d multimodal \\
Unique 3-tuple SIDs
& 11,615 & 11,649 & 17,361 & 127,844 & 153,046 \\
Collision rate
& 2.6\% & 3.7\% & 5.4\% & 9.5\% & 6.5\% \\
\bottomrule
\end{tabular}
\end{table*}

For Sports, Beauty, and Toys, the unique three-level SID rates are 94.6\%, 96.3\%, and 97.4\%, respectively, corresponding to collision rates of 5.4\%, 3.7\%, and 2.6\%. The maximum deduplication indices observed are 13, 35, and 11. All 256 codes are utilized at each quantization level. Fashion-140k has a higher collision rate because its items are quantized using a codebook trained on the larger 799,953-item pool rather than directly on the subset. The deduplication token therefore provides a deterministic mechanism for distinguishing items that share the same three-level semantic code.

\subsubsection{Training-Data Construction}
\label{sec:supp_training_data}

\paragraph{Hierarchical-loss training.}
The hierarchical loss is a training objective rather than a separate data-generation pipeline. It is applied to the same semantic-ID conversational training corpus used by the flat causal CE baseline, reweighting the per-level losses without changing the underlying examples.

For Sports, Beauty, and Toys, conversational SFT data are generated from item metadata, semantic IDs, and user sequences using fixed instruction templates. Sports and Beauty use 26 conversation types, while Toys uses 22 because several types requiring lower-coverage fields are omitted. Repeated instructions mapping to the same SID are intentionally retained to reinforce the SID-item mapping and provide surface-form diversity. Examples are shuffled with a fixed seed and split 92/5/3 into training and validation.

The conversation types fall into five groups:

\begin{itemize}
\item \textbf{SID$\rightarrow$text decoding:} recover title, description, context, category, or brand from a semantic ID.
\item \textbf{Text$\rightarrow$SID encoding:} predict a semantic ID from title, description, or brand information.
\item \textbf{Sequential next-item prediction:} predict the next item from histories of length 2, 3, 5 and all. 
\item \textbf{Co-engagement:} predict co-purchased, frequently bought, or similar products.
\item \textbf{Category/prefix structure understanding:} reason over shared semantic-ID prefixes and their associated categories or examples.
\end{itemize}

This produces 2,383,517, 1,634,627, and 1,320,505 training conversations for Sports, Beauty, and Toys, respectively.

\paragraph{Attribute-loss training.}
The attribute-conditioned loss is evaluated on Fashion-140k and Internal-fashion, where structured attributes are sufficiently dense. The loss is applied as an auxiliary objective over the training corpus rather than being restricted exclusively to attribute-conditioned conversations.

\textbf{Fashion-140k.}
The 1,336,658-example training corpus is organized into five families:
\begin{itemize}
\item \textbf{AF1 -- SID$\leftrightarrow$text/attribute mappings:} approximately 700K examples covering SID-to-text, text-to-SID, and attribute-to-SID mappings.
\item \textbf{AF2 -- attribute-based retrieval:} approximately 27K examples using gender, product type, color, style, occasion, brand, and material, including single-, two-, and three-attribute queries.
\item \textbf{AF3 -- sequential prediction:} approximately 200K next-item and co-purchase examples.
\item \textbf{AF4 -- outfit color coordination:} approximately 1K examples derived from Fashionpedia and Polyvore color-compatibility rules.
\item \textbf{AF5 -- semantic hierarchy understanding:} approximately 117K examples based on SID-prefix clusters and similar-item relationships.
\end{itemize}

AF2 and AF4 provide the most direct attribute supervision because their targets are explicitly organized by shared attributes such as color, material, gender, and style.

\textbf{Internal-fashion.}
The Internal-fashion SFT corpus contains 368,482 training and 19,482 validation conversations using the same 57-type schema as the underlying pipeline. In addition to SID$\leftrightarrow$text tasks, it contains exhaustive attribute-conjunction retrieval examples over combinations of brand, color, material, gender, and category. Examples include queries for all products matching a particular color-category or brand-color combination.

Rather than restricting the auxiliary objective to these explicitly color-named examples, the color loss is applied as a soft per-example reweighting over the entire corpus. Examples whose conversation type contains ``color'' receive the full auxiliary weight ($\alpha=0.1$), while all other examples receive a reduced scale of $0.2\alpha$. Thus, even items that do not occur in explicitly color-conditioned conversations receive a weaker color-anchoring gradient.

The color objective uses soft-label cross-entropy over color-similar candidates, with similarity determined by $\exp(-\Delta E/T)$ in CIE-LAB~\cite{CIE1976Lab} color space and an additional frequency-based weighting term that upweights rare items. This formulation allows attribute supervision to provide stronger fine-grained signals while retaining the broad conversational training distribution.

\subsection{Implementation Details}

\subsubsection{Hierarchical Loss}
\label{sec:supp_hloss_impl}

We provide implementation details for the confidence-gated hierarchical loss used to evaluate Sports, Beauty, and Toys. The loss combines static level weighting, confidence-based gating, annealed blending, and scheduled sampling.

\paragraph{Static level weighting.}
Each SID level $l$ is assigned a fixed base weight $w_l$, with larger weights for coarser levels. Non-SID tokens retain a weight of $1.0$. This provides hierarchical supervision from the beginning of training, independently of the model's confidence.

\paragraph{Confidence-based gating.}
For each SID level $l>0$, the base weight is modulated by the model's confidence in the correct codes at all preceding levels. Specifically, let $p_{\mathrm{correct}}(s_{i,k})$ denote the predicted probability of the ground-truth code $s_{i,k}$ at level $k$. The effective weight is
\begin{equation}
w_l^{\mathrm{eff}}
=
w_l
\left[
\alpha
+
(1-\alpha)
\prod_{k<l}
p_{\mathrm{correct}}(s_{i,k})^{\gamma}
\right],
\end{equation}
where
\begin{equation}
p_{\mathrm{correct}}(s_{i,k})
\leftarrow
\operatorname{clamp}
\left(
p_{\mathrm{correct}}(s_{i,k}),
\mathrm{min}=\mathrm{floor}
\right).
\end{equation}
The confidence values are detached from the computation graph, preventing the model from increasing the gate through the gating pathway itself. We use a gating floor of $0.2$ and $\gamma=1.0$.

\begin{table}[h]
\centering
\caption{Hyperparameter settings for the hierarchical loss. Base weights are ordered from the coarsest to the finest SID level. The same configuration is used across Sports, Beauty, and Toys.}
\small
\begin{tabular}{lc}
\toprule
\textbf{Hyperparameter} & \textbf{Setting} \\
\midrule
Base weights $(w_0,w_1,w_2,w_3)$ & $(4.0,3.0,2.0,1.0)$ \\
Gating floor & 0.2 \\
Gating sharpness $\gamma$ & 1.0 \\
Max scheduled-sampling rate & 0.3 \\
Scheduled-sampling anneal steps & 200{,}000 \\
Blend anneal steps & 300{,}000 \\
\midrule
Learning rate & $2\times10^{-5}$ \\
Per-device batch size & 4 \\
Gradient accumulation & 8 (effective batch size 32) \\
Epochs & 20 \\
LR schedule & Cosine, 3\% linear warmup \\
Weight decay & 0.01 \\
Gradient clip norm & 1.0 \\
Optimizer & \texttt{adamw\_8bit} \\
Maximum sequence length & 2048 tokens \\
\bottomrule
\end{tabular}
\label{tab:supp_hloss_hparams}
\end{table}

\begin{table*}[h]
\centering
\caption{One example per conversation sub-type in the Toys dataset, grouped into the five families of Section~\ref{sec:supp_training_data}. Semantic IDs are abbreviated as \{sid\_W,X,Y,Z\} in this table only, where $W$, $X$, $Y$, and $Z$ denote the four semantic-ID levels.}
\small
\renewcommand{\arraystretch}{1.3}
\begin{tabular}{@{}p{2.6cm}p{2.4cm}p{5.1cm}p{4.6cm}@{}}
\toprule
\textbf{Group} & \textbf{Sub-type ($n$)} & \textbf{User turn} & \textbf{Assistant turn} \\
\midrule

\multirow{5}{*}{\shortstack[l]{SID $\to$ text\\(decoding)}}
& sid\_to\_title (11{,}865)
& Product \{sid\_58,401,722,768\} has title:
& Doctor Who -- Master Laser Screwdriver \\

& sid\_to\_description (10{,}736)
& Item \{sid\_30,341,570,768\} is described as:
& Only Playhut structures provide children with hours of interactive adventure. EZ Twist technology allows for instant setup and quick fold-down\ldots \\

& sid\_to\_context (11{,}865)
& Tell me about product \{sid\_107,383,713,768\}:
& Title: Crayola Tub of Fun; Description: Crayola Tub of Fun; Category: Toys \& Games \\

& sid\_to\_category (11{,}924)
& Product \{sid\_122,381,749,768\} belongs to category:
& Toys \& Games $>$ Games $>$ Board Games \\

& sid\_to\_brand (10{,}199)
& Product \{sid\_77,477,593,768\} is made by:
& Parker Brothers \\

\midrule

\multirow{3}{*}{\shortstack[l]{Text $\to$ SID\\(encoding)}}
& title\_to\_sid (11{,}865)
& The product ``Bandai Tamashii Stage Act.5 for Mechanics Clear'' has SemanticID:
& \{sid\_56,289,628,768\} \\

& description\_to\_sid (7{,}695)
& A product described as ``Hit the road in style---it's the coolest way to travel and soak up some sun. This two-seater convertible has sweet details\ldots'' has SemanticID:
& \{sid\_176,504,546,768\} \\

& brand\_to\_sid (10{,}199)
& A Toys \& Games product made by ``Learning Curve'' has SemanticID:
& \{sid\_199,352,680,768\} \\

\midrule

\multirow{3}{*}{\shortstack[l]{Sequential\\next-item\\prediction}}
& seq\_last\_2 (128{,}773)
& User's last purchases: \{sid\_94,452,645,768\}, \{sid\_246,397,725,768\}. Next:
& \{sid\_49,283,602,768\} \\

& seq\_last\_3 (109{,}361)
& Based on recent purchases: \{sid\_53,484,691,768\}, \{sid\_83,423,723,768\}, \{sid\_211,471,616,768\}, next item:
& \{sid\_12,442,677,768\} \\

& seq\_last\_5 (70{,}537)
& Purchase sequence: \{sid\_129,456,766,768\}, \{sid\_6,387,739,768\}, \{sid\_6,302,559,768\}, \{sid\_6,387,591,768\}, \{sid\_79,487,764,768\}. Predict next:
& \{sid\_6,387,554,768\} \\

\midrule

\multirow{4}{*}{\shortstack[l]{Co-engagement}}
& copurchase\_forward (474{,}246)
& A user who bought \{sid\_167,411,542,768\} (Melissa \& Doug Sunny Patch Happy Giddy Chair) might also buy:
& \{sid\_210,329,533,768\} \\

& copurchase\_backward (472{,}066)
& Before buying \{sid\_192,300,739,768\} (Playskool Learnimals ABC Adventure Elefun Toy), users often buy:
& \{sid\_68,326,561,768\} \\

& similar\_products (4{,}403)
& Products similar to \{sid\_19,273,603,768\} include:
& \{sid\_19,273,661,768\} \\

\midrule

\multirow{4}{*}{\shortstack[l]{Category /\\prefix structure\\understanding}}
& prefix\_category (256)
& Products starting with \{sid\_118\} are typically:
& Toys \& Games products \\

& prefix\_2level\_category (442)
& Products starting with \{sid\_140,261\} are:
& Toys \& Games items \\

& prefix\_examples (256)
& Examples of products with prefix \{sid\_20\}:
& \{sid\_20,374,673,768\}, \{sid\_20,491,661,768\}, \{sid\_20,354,723,768\} \\

& category\_transition (8)
& After buying from Health \& Personal Care, users often buy from:
& Toys \& Games \\

\bottomrule
\end{tabular}
\label{tab:supp_conv_examples_toys}
\end{table*}

\paragraph{Annealed confidence blending.}
The coefficient $\alpha$ controls the transition from static weighting to confidence-based weighting. It is initialized to $1.0$, corresponding to purely static weighting, and linearly decreases to $0.0$ over $300{,}000$ optimizer steps. Thus, training initially emphasizes a stable hierarchical curriculum and gradually transitions to adaptive supervision based on the model's confidence in preceding levels.

\paragraph{Scheduled sampling.}
We additionally use scheduled sampling to reduce the exposure gap between teacher-forced training and autoregressive inference. A first forward pass, performed without gradient computation, samples the predicted code at the coarsest SID level from the model's categorical distribution. With probability $\mathtt{ss\_prob}$, this prediction replaces the corresponding ground-truth code in the input sequence. A second forward pass then computes the hierarchical loss on the resulting sequence. The scheduled-sampling probability is linearly increased from $0$ to $0.3$ over $200{,}000$ optimizer steps. Multinomial sampling is used rather than greedy decoding to expose the model to the stochasticity of autoregressive generation.

\paragraph{Category-specific configurations.}
The hierarchical-loss mechanism is shared across all three datasets, with category-specific differences in the number of SID levels and their base weights. Sports, Beauty and Toys use four SID codes with weights $(4.0,3.0,2.0,1.0)$. The fourth code corresponds to the deduplication code in addition to the three residual-quantization levels. All other hierarchical-loss hyperparameters are shared across datasets, as summarized in Table~\ref{tab:supp_hloss_hparams}.

\paragraph{Training and inference.}
Training is performed in two stages. First, only the newly introduced SID embeddings are optimized for $50{,}000$ steps using flat cross-entropy with a learning rate of $1\times10^{-3}$. The resulting checkpoint initializes the second stage, in which all model parameters are jointly fine-tuned using the hierarchical loss and the configuration in Table~\ref{tab:supp_hloss_hparams}.

During evaluation, autoregressive generation is constrained to catalog-valid SID sequences using a prefix-tree logits processor constructed from up to $100$K training sequences. If a generated prefix is not present in the tree, decoding falls back to a per-position vocabulary mask over the 256 valid SID codes at the corresponding level.

\subsubsection{Attribute-Specific Losses}
\label{sec:supp_attribute_impl}

\begin{table*}[h]
\centering
\caption{Hyperparameter settings for the hierarchical-loss component ablation. Base weights are ordered from the coarsest to the finest SID level. All variants use three SID levels.}
\small
\begin{tabular}{lccccc}
\toprule
\textbf{Variant} & \textbf{Base weights $(w_1,w_2,w_3,w_4)$} 
& \textbf{$\alpha$ schedule} & \textbf{Gate floor / $\gamma$} 
& \textbf{$\mathtt{ss\_max\_prob}$} \\
\midrule
CE & $(1.0,1.0,1.0,1.0)$ 
& $\alpha \equiv 1.0$ & -- / -- & $0$ \\
HL w/ SW & $(4.0,3.0,2.0,1.0)$ 
& $\alpha \equiv 1.0$ & -- / -- & $0$ \\
HL w/ CG & $(1.0,1.0,1.0,1.0)$ 
& $1.0 \rightarrow 0.0$ over $300{,}000$ steps 
& $0.2$ / $1.0$ & $0$ \\
HL w/ SS & $(1.0,1.0,1.0,1.0)$ 
& $\alpha \equiv 1.0$ & -- / -- & $0.3$ \\
HL w/ SW, CG, SS (full) & $(4.0,3.0,2.0,1.0)$ 
& $1.0 \rightarrow 0.0$ over $300{,}000$ steps 
& $0.2$ / $1.0$ & $0.3$ \\
\bottomrule
\end{tabular}
\label{tab:supp_hloss_ablation_hparams}
\end{table*}

\paragraph{Shared architecture and attachment.}
We use the same auxiliary-head architecture for both color and material composition prediction. The head is attached to the last-layer hidden state at the L2 semantic-ID prediction position, $h_{L2}$, after the model has consumed the preceding L0 and L1 codes. Each head is a two-layer MLP,
\[
\operatorname{MLP}(h)
=
\operatorname{Linear}(4096,64)
\rightarrow
\operatorname{GELU}
\rightarrow
\operatorname{Linear}(64,d_{\mathrm{out}}),
\]
where $d_{\mathrm{out}}=3$ for color and $d_{\mathrm{out}}=15$ for material composition. The heads are trained jointly with the full LLM backbone. The overall objective is
\[
\mathcal{L}
=
\mathcal{L}_{\mathrm{CE}}
+
\lambda_{\mathrm{attr}}\mathcal{L}_{\mathrm{Huber}},
\]
where the attribute loss is applied at every SID position in every conversation, without conditioning on conversation type.

\paragraph{Color supervision.}
Color targets are derived from catalog-level CIE-LAB annotations. For each SID sequence $(L0,L1,L2,L3)$, we average the LAB values of all catalog items assigned to that triple. Before computing the loss, $L^*$ is normalized as $(L^*-50)/50$, while $a^*$ and $b^*$ are normalized as $a^*/80$ and $b^*/80$. We use a mean-reduced Huber loss with $\delta=0.1$:
\[
\mathcal{L}_{\mathrm{color}}
=
\operatorname{Huber}_{0.1}
\left(
\hat{\mathbf{y}}_{\mathrm{color}},
\mathbf{y}_{\mathrm{color}}
\right).
\]
To emphasize data-sparse semantic IDs, each example is weighted according to its L2-code frequency:
\[
w_i
=
\operatorname{clip}
\left(
\left(
\frac{f_{\mathrm{med}}}{f_i}
\right)^{0.5},
1,5
\right).
\]
The weighted loss is multiplied by $\lambda_{\mathrm{color}}=0.1$.

\paragraph{Material-composition supervision.}
Material annotations are converted into 15-dimensional composition vectors comprising 14 canonical material categories plus an \textit{other} category. Explicit percentages are retained, while unspecified components divide the remaining mass uniformly; the resulting vector is L1-normalized to sum to one. SID-level targets are obtained by averaging composition vectors across all items sharing the same SID triple. We apply a mean-reduced Huber loss with $\delta=0.1$ directly in composition space:
\[
\mathcal{L}_{\mathrm{material}}
=
\operatorname{Huber}_{0.1}
\left(
\hat{\mathbf{y}}_{\mathrm{material}},
\mathbf{y}_{\mathrm{material}}
\right).
\]
Unlike color, material supervision uses no frequency-based tail weighting. We set $\lambda_{\mathrm{material}}=1.0$ to retain sufficient gradient magnitude for the smaller raw composition loss.

\paragraph{Training details.}
Both auxiliary heads are dynamically attached to the pretrained model and optimized jointly with the full backbone. The attribute loss is evaluated for every SID-bearing training example rather than only attribute-specific conversations. For these experiments, the CE component is computed over the full input sequence (prompt and completion), corresponding to \texttt{completion\_only\\\_loss=False}. The auxiliary heads are saved separately from the Hugging Face model checkpoint.

\subsubsection{Hyperparameter Settings for the Hierarchical-Loss Component Ablation}
\label{sec:supp_hloss_ablation}

Table~\ref{tab:supp_hloss_ablation_hparams} reports the hyperparameter settings used for the hierarchical-loss component ablation. Each variant is derived from the full configuration by disabling the other mechanisms, allowing the contribution of static weighting (SW), confidence gating (CG), and scheduled sampling (SS) to be isolated. The \textbf{CE} baseline disables all three mechanisms and applies uniform weighting to all SID tokens. \textbf{HL w/ SW} uses static level weights while disabling confidence gating and scheduled sampling. \textbf{HL w/ CG} uses confidence-based gating with uniform base weights, while disabling scheduled sampling. \textbf{HL w/ SS} uses scheduled sampling with uniform base weights and no confidence gating. \textbf{HL w/ SW, CG, SS} corresponds to the full hierarchical-loss configuration.

All variants use the same optimizer and training configuration: learning rate $2\times10^{-5}$, effective batch size 16, 20 training epochs, and a cosine learning-rate schedule. Scheduled sampling, when enabled, uses an annealing period of $200{,}000$ steps. The static level weights are $(3.0,2.0,1.0)$ from the coarsest to the finest level. Confidence gating uses a floor of $0.2$, $\gamma=1.0$, and an $\alpha$ schedule that linearly decreases from $1.0$ to $0.0$ over $300{,}000$ steps.

\subsubsection{Conversation Type Examples (Toys)}
\label{sec:supp_conversation_examples}

Table~\ref{tab:supp_conv_examples_toys} gives one representative example of every conversation
sub-type in the Toys dataset ($n=1{,}390{,}005$ total conversations), grouped into the five
families described above. Semantic IDs are shown in the
model's native token format, \texttt{<|sid\_start|><|sid\_W|>\\<|sid\_X|><|sid\_Y|><|sid\_Z|><|sid\_end|>},
where $W, X, Y, Z \in [0, 255]$ index the four RQ-VAE codebook levels.


\bibliographystyle{ACM-Reference-Format}
\bibliography{sample-base.bib}

@String{Computing = "Computing" }

@String{Computer = "{IEEE} Computer" }

@String{Springer = "Springer-Verlag" }

@ArtifactSoftware{R,
    title = {R: A Language and Environment for Statistical Computing},
    author = {{R Core Team}},
    organization = {R Foundation for Statistical Computing},
    address = {Vienna, Austria},
    year = {2019},
    url = {https://www.R-project.org/},
}

@book{CIE1976Lab,
  author    = {{CIE}},
  title     = {Colorimetry -- Part 4: {CIE} 1976 $L^*a^*b^*$ Colour Space},
  publisher = {Commission Internationale de l'Eclairage},
  address   = {Vienna, Austria},
  number    = {ISO 11664-4:2008(E) / CIE S 014-4/E:2007},
  year      = {2007}
}

@article{hou2024bridging,
  title={Bridging Language and Items for Retrieval and Recommendation},
  author={Hou, Yupeng and Li, Jiacheng and He, Zhankui and Yan, An and Chen, Xiusi and McAuley, Julian},
  journal={arXiv preprint arXiv:2403.03952},
  year={2024}
}

@misc{sun2019bert4recsequentialrecommendationbidirectional,
      title={BERT4Rec: Sequential Recommendation with Bidirectional Encoder Representations from Transformer}, 
      author={Fei Sun and Jun Liu and Jian Wu and Changhua Pei and Xiao Lin and Wenwu Ou and Peng Jiang},
      year={2019},
      eprint={1904.06690},
      archivePrefix={arXiv},
      primaryClass={cs.IR},
      url={https://arxiv.org/abs/1904.06690}, 
}

@misc{kang2018selfattentivesequentialrecommendation,
      title={Self-Attentive Sequential Recommendation}, 
      author={Wang-Cheng Kang and Julian McAuley},
      year={2018},
      eprint={1808.09781},
      archivePrefix={arXiv},
      primaryClass={cs.IR},
      url={https://arxiv.org/abs/1808.09781}, 
}

@misc{tan2024idgenrecllmrecsysalignmenttextual,
      title={IDGenRec: LLM-RecSys Alignment with Textual ID Learning}, 
      author={Juntao Tan and Shuyuan Xu and Wenyue Hua and Yingqiang Ge and Zelong Li and Yongfeng Zhang},
      year={2024},
      eprint={2403.19021},
      archivePrefix={arXiv},
      primaryClass={cs.IR},
      url={https://arxiv.org/abs/2403.19021}, 
}

@misc{liu2025onerecthinkintextreasoninggenerative,
      title={OneRec-Think: In-Text Reasoning for Generative Recommendation}, 
      author={Zhanyu Liu and Shiyao Wang and Xingmei Wang and Rongzhou Zhang and Jiaxin Deng and Honghui Bao and Jinghao Zhang and Wuchao Li and Pengfei Zheng and Xiangyu Wu and Yifei Hu and Qigen Hu and Xinchen Luo and Lejian Ren and Zixing Zhang and Qianqian Wang and Kuo Cai and Yunfan Wu and Hongtao Cheng and Zexuan Cheng and Lu Ren and Huanjie Wang and Yi Su and Ruiming Tang and Kun Gai and Guorui Zhou},
      year={2025},
      eprint={2510.11639},
      archivePrefix={arXiv},
      primaryClass={cs.IR},
      url={https://arxiv.org/abs/2510.11639}, 
}

@misc{rajput2023recommendersystemsgenerativeretrieval,
      title={Recommender Systems with Generative Retrieval}, 
      author={Shashank Rajput and Nikhil Mehta and Anima Singh and Raghunandan H. Keshavan and Trung Vu and Lukasz Heldt and Lichan Hong and Yi Tay and Vinh Q. Tran and Jonah Samost and Maciej Kula and Ed H. Chi and Maheswaran Sathiamoorthy},
      year={2023},
      eprint={2305.05065},
      archivePrefix={arXiv},
      primaryClass={cs.IR},
      url={https://arxiv.org/abs/2305.05065}, 
}

@misc{zhai2024actionsspeaklouderwords,
      title={Actions Speak Louder than Words: Trillion-Parameter Sequential Transducers for Generative Recommendations}, 
      author={Jiaqi Zhai and Lucy Liao and Xing Liu and Yueming Wang and Rui Li and Xuan Cao and Leon Gao and Zhaojie Gong and Fangda Gu and Michael He and Yinghai Lu and Yu Shi},
      year={2024},
      eprint={2402.17152},
      archivePrefix={arXiv},
      primaryClass={cs.LG},
      url={https://arxiv.org/abs/2402.17152}, 
}

@misc{zeghidour2021soundstreamendtoendneuralaudio,
      title={SoundStream: An End-to-End Neural Audio Codec}, 
      author={Neil Zeghidour and Alejandro Luebs and Ahmed Omran and Jan Skoglund and Marco Tagliasacchi},
      year={2021},
      eprint={2107.03312},
      archivePrefix={arXiv},
      primaryClass={cs.SD},
      url={https://arxiv.org/abs/2107.03312}, 
}

@article{deng2025onerec,
  title={Onerec: Unifying retrieve and rank with generative recommender and iterative preference alignment},
  author={Deng, Jiaxin and Wang, Shiyao and Cai, Kuo and Ren, Lejian and Hu, Qigen and Ding, Weifeng and Luo, Qiang and Zhou, Guorui},
  journal={arXiv preprint arXiv:2502.18965},
  year={2025}
}

@inproceedings{lee2022rqvae,
  title={Autoregressive image generation using residual quantization},
  author={Lee, Doyup and Kim, Chiheon and Kim, Saehoon and Cho, Minsu and Han, Wook-Shin},
  booktitle={2022 IEEE/CVF conference on computer vision and pattern recognition (CVPR)},
  pages={11513--11522},
  year={2022},
  organization={IEEE}
}

@inproceedings{luo2024qarm,
  title={Qarm: Quantitative alignment multi-modal recommendation at kuaishou},
  author={Luo, Xinchen and Cao, Jiangxia and Sun, Tianyu and Yu, Jinkai and Huang, Rui and Yuan, Wei and Lin, Hezheng and Zheng, Yichen and Wang, Shiyao and Hu, Qigen and others},
  booktitle={Proceedings of the 34th ACM International Conference on Information and Knowledge Management},
  pages={5915--5922},
  year={2025}
}

@inproceedings{covington2016youtube,
author = {Covington, Paul and Adams, Jay and Sargin, Emre},
title = {Deep Neural Networks for YouTube Recommendations},
year = {2016},
isbn = {9781450340359},
publisher = {Association for Computing Machinery},
address = {New York, NY, USA},
url = {https://doi.org/10.1145/2959100.2959190},
doi = {10.1145/2959100.2959190},
booktitle = {Proceedings of the 10th ACM Conference on Recommender Systems},
pages = {191–198},
numpages = {8},
location = {Boston, Massachusetts, USA},
series = {RecSys '16}
}

@article{wang2020cold,
  title={Cold: Towards the next generation of pre-ranking system},
  author={Wang, Zhe and Zhao, Liqin and Jiang, Biye and Zhou, Guorui and Zhu, Xiaoqiang and Gai, Kun},
  journal={arXiv preprint arXiv:2007.16122},
  year={2020}
}

@inproceedings{pei2019prm,
author = {Pei, Changhua and Zhang, Yi and Zhang, Yongfeng and Sun, Fei and Lin, Xiao and Sun, Hanxiao and Wu, Jian and Jiang, Peng and Ge, Junfeng and Ou, Wenwu and Pei, Dan},
title = {Personalized re-ranking for recommendation},
year = {2019},
isbn = {9781450362436},
publisher = {Association for Computing Machinery},
address = {New York, NY, USA},
url = {https://doi.org/10.1145/3298689.3347000},
doi = {10.1145/3298689.3347000},
booktitle = {Proceedings of the 13th ACM Conference on Recommender Systems},
pages = {3–11},
numpages = {9},
location = {Copenhagen, Denmark},
series = {RecSys '19}
}

@inproceedings{ma2021pre,
author = {Ma, Xu and Wang, Pengjie and Zhao, Hui and Liu, Shaoguo and Zhao, Chuhan and Lin, Wei and Lee, Kuang-Chih and Xu, Jian and Zheng, Bo},
title = {Towards a Better Tradeoff between Effectiveness and Efficiency in Pre-Ranking: A Learnable Feature Selection based Approach},
year = {2021},
isbn = {9781450380379},
publisher = {Association for Computing Machinery},
address = {New York, NY, USA},
url = {https://doi.org/10.1145/3404835.3462979},
doi = {10.1145/3404835.3462979},
booktitle = {Proceedings of the 44th International ACM SIGIR Conference on Research and Development in Information Retrieval},
pages = {2036–2040},
numpages = {5},
location = {Virtual Event, Canada},
series = {SIGIR '21}
}

@inproceedings{song2022rethinking,
author = {Song, Jinbo and Huang, Ruoran and Wang, Xinyang and Huang, Wei and Yu, Qian and Chen, Mingming and Yao, Yafei and Fan, Chaosheng and Peng, Changping and Lin, Zhangang and Hu, Jinghe and Shao, Jingping},
title = {Rethinking Large-scale Pre-ranking System: Entire-chain Cross-domain Models},
year = {2022},
isbn = {9781450392365},
publisher = {Association for Computing Machinery},
address = {New York, NY, USA},
url = {https://doi.org/10.1145/3511808.3557683},
doi = {10.1145/3511808.3557683},
booktitle = {Proceedings of the 31st ACM International Conference on Information \& Knowledge Management},
pages = {4495–4499},
numpages = {5},
location = {Atlanta, GA, USA},
series = {CIKM '22}
}

@inproceedings{wang2024letter,
author = {Wang, Wenjie and Bao, Honghui and Lin, Xinyu and Zhang, Jizhi and Li, Yongqi and Feng, Fuli and Ng, See-Kiong and Chua, Tat-Seng},
title = {Learnable Item Tokenization for Generative Recommendation},
year = {2024},
isbn = {9798400704369},
publisher = {Association for Computing Machinery},
address = {New York, NY, USA},
url = {https://doi.org/10.1145/3627673.3679569},
doi = {10.1145/3627673.3679569},
booktitle = {Proceedings of the 33rd ACM International Conference on Information and Knowledge Management},
pages = {2400–2409},
numpages = {10},
location = {Boise, ID, USA},
series = {CIKM '24}
}

@inproceedings{wei2021contrastive,
  title={Contrastive learning for cold-start recommendation},
  author={Wei, Yinwei and Wang, Xiang and Li, Qi and Nie, Liqiang and Li, Yan and Li, Xuanping and Chua, Tat-Seng},
  booktitle={Proceedings of the 29th ACM international conference on multimedia},
  pages={5382--5390},
  year={2021}
}

@article{hidasi2016gru4rec,
  title={Session-based recommendations with recurrent neural networks},
  author={Hidasi, Bal{\'a}zs and Karatzoglou, Alexandros and Baltrunas, Linas and Tikk, Domonkos},
  journal={arXiv preprint arXiv:1511.06939},
  year={2015}
}

@article{tay2022dsi,
  title={Transformer memory as a differentiable search index},
  author={Tay, Yi and Tran, Vinh and Dehghani, Mostafa and Ni, Jianmo and Bahri, Dara and Mehta, Harsh and Qin, Zhen and Hui, Kai and Zhao, Zhe and Gupta, Jai and others},
  journal={Advances in neural information processing systems},
  volume={35},
  pages={21831--21843},
  year={2022}
}

@article{wang2022nci,
  title={A neural corpus indexer for document retrieval},
  author={Wang, Yujing and Hou, Yingyan and Wang, Haonan and Miao, Ziming and Wu, Shibin and Chen, Qi and Xia, Yuqing and Chi, Chengmin and Zhao, Guoshuai and Liu, Zheng and others},
  journal={Advances in Neural Information Processing Systems},
  volume={35},
  pages={25600--25614},
  year={2022}
}

@inproceedings{wang2024eager,
author = {Wang, Ye and Xun, Jiahao and Hong, Minjie and Zhu, Jieming and Jin, Tao and Lin, Wang and Li, Haoyuan and Li, Linjun and Xia, Yan and Zhao, Zhou and Dong, Zhenhua},
title = {EAGER: Two-Stream Generative Recommender with Behavior-Semantic Collaboration},
year = {2024},
isbn = {9798400704901},
publisher = {Association for Computing Machinery},
address = {New York, NY, USA},
url = {https://doi.org/10.1145/3637528.3671775},
doi = {10.1145/3637528.3671775},
booktitle = {Proceedings of the 30th ACM SIGKDD Conference on Knowledge Discovery and Data Mining},
pages = {3245–3254},
numpages = {10},
location = {Barcelona, Spain},
series = {KDD '24}
}

@article{liu2024mmgrec,
  title={Mmgrec: Multimodal generative recommendation with transformer model},
  author={Liu, Han and Wei, Yinwei and Song, Xuemeng and Guan, Weili and Li, Yuan-Fang and Nie, Liqiang},
  journal={ACM Transactions on Multimedia Computing, Communications and Applications},
  year={2024},
  publisher={ACM New York, NY}
}

@article{li2023gpt4rec,
  title={GPT4Rec: A generative framework for personalized recommendation and user interests interpretation},
  author={Li, Jinming and Zhang, Wentao and Wang, Tian and Xiong, Guanglei and Lu, Alan and Medioni, Gerard},
  journal={arXiv preprint arXiv:2304.03879},
  year={2023}
}

@article{gao2023chatrec,
  title={Chat-rec: Towards interactive and explainable llms-augmented recommender system},
  author={Gao, Yunfan and Sheng, Tao and Xiang, Youlin and Xiong, Yun and Wang, Haofen and Zhang, Jiawei},
  journal={arXiv preprint arXiv:2303.14524},
  year={2023}
}

@inproceedings{hou2024zeroshot,
  title={Large language models are zero-shot rankers for recommender systems},
  author={Hou, Yupeng and Zhang, Junjie and Lin, Zihan and Lu, Hongyu and Xie, Ruobing and McAuley, Julian and Zhao, Wayne Xin},
  booktitle={European conference on information retrieval},
  pages={364--381},
  year={2024},
  organization={Springer}
}

@inproceedings{geng2022p5,
  title={Recommendation as language processing (rlp): A unified pretrain, personalized prompt \& predict paradigm (p5)},
  author={Geng, Shijie and Liu, Shuchang and Fu, Zuohui and Ge, Yingqiang and Zhang, Yongfeng},
  booktitle={Proceedings of the 16th ACM conference on recommender systems},
  pages={299--315},
  year={2022}
}

@inproceedings{zheng2024lcrec,
  title={Adapting large language models by integrating collaborative semantics for recommendation},
  author={Zheng, Bowen and Hou, Yupeng and Lu, Hongyu and Chen, Yu and Zhao, Wayne Xin and Chen, Ming and Wen, Ji-Rong},
  booktitle={2024 IEEE 40th International Conference on Data Engineering (ICDE)},
  pages={1435--1448},
  year={2024},
  organization={IEEE}
}

@inproceedings{morin2005hsoftmax,
  title = 	 {Hierarchical Probabilistic Neural Network Language Model},
  author =       {Morin, Frederic and Bengio, Yoshua},
  booktitle = 	 {Proceedings of the Tenth International Workshop on Artificial Intelligence and Statistics},
  pages = 	 {246--252},
  year = 	 {2005},
  editor = 	 {Cowell, Robert G. and Ghahramani, Zoubin},
  volume = 	 {R5},
  series = 	 {Proceedings of Machine Learning Research},
  month = 	 {06--08 Jan},
  publisher =    {PMLR},
  url = 	 {https://proceedings.mlr.press/r5/morin05a.html},
  note =         {Reissued by PMLR on 30 March 2021.}
}

@inproceedings{bertinetto2020hierarchy,
  title={Making better mistakes: Leveraging class hierarchies with deep networks},
  author={Bertinetto, Luca and Mueller, Romain and Tertikas, Konstantinos and Samangooei, Sina and Lord, Nicholas A},
  booktitle={2020 IEEE/CVF Conference on Computer Vision and Pattern Recognition (CVPR)},
  pages={12503--12512},
  year={2020},
  organization={IEEE}
}

@InProceedings{lee2015deeply,
  title = 	 {{Deeply-Supervised Nets}},
  author = 	 {Lee, Chen-Yu and Xie, Saining and Gallagher, Patrick and Zhang, Zhengyou and Tu, Zhuowen},
  booktitle = 	 {Proceedings of the Eighteenth International Conference on Artificial Intelligence and Statistics},
  pages = 	 {562--570},
  year = 	 {2015},
  editor = 	 {Lebanon, Guy and Vishwanathan, S. V. N.},
  volume = 	 {38},
  series = 	 {Proceedings of Machine Learning Research},
  address = 	 {San Diego, California, USA},
  month = 	 {09--12 May},
  publisher =    {PMLR},
  url = 	 {https://proceedings.mlr.press/v38/lee15a.html}
}

@article{stretcu2021coarsetofine,
  title   = {Coarse-to-Fine Curriculum Learning},
  author  = {Stretcu, Otilia and Platanios, Emmanouil Antonios and Mitchell, Tom M. and P{\'o}czos, Barnab{\'a}s},
  journal = {arXiv preprint arXiv:2106.04072},
  year    = {2021}
}

@inproceedings{bengio2015scheduledsampling,
author = {Bengio, Samy and Vinyals, Oriol and Jaitly, Navdeep and Shazeer, Noam},
title = {Scheduled sampling for sequence prediction with recurrent Neural networks},
year = {2015},
publisher = {MIT Press},
address = {Cambridge, MA, USA},
booktitle = {Proceedings of the 29th International Conference on Neural Information Processing Systems - Volume 1},
pages = {1171–1179},
numpages = {9},
location = {Montreal, Canada},
series = {NIPS'15}
}

@inproceedings{ma2018mmoe,
author = {Ma, Jiaqi and Zhao, Zhe and Yi, Xinyang and Chen, Jilin and Hong, Lichan and Chi, Ed H.},
title = {Modeling Task Relationships in Multi-task Learning with Multi-gate Mixture-of-Experts},
year = {2018},
isbn = {9781450355520},
publisher = {Association for Computing Machinery},
address = {New York, NY, USA},
url = {https://doi.org/10.1145/3219819.3220007},
doi = {10.1145/3219819.3220007},
booktitle = {Proceedings of the 24th ACM SIGKDD International Conference on Knowledge Discovery \& Data Mining},
pages = {1930–1939},
numpages = {10},
location = {London, United Kingdom},
series = {KDD '18}
}

@article{brown2020gpt3,
  title={Language models are few-shot learners},
  author={Brown, Tom and Mann, Benjamin and Ryder, Nick and Subbiah, Melanie and Kaplan, Jared D and Dhariwal, Prafulla and Neelakantan, Arvind and Shyam, Pranav and Sastry, Girish and Askell, Amanda and others},
  journal={Advances in neural information processing systems},
  volume={33},
  pages={1877--1901},
  year={2020}
}

@article{touvron2023llama,
  title={Llama: Open and efficient foundation language models},
  author={Touvron, Hugo and Lavril, Thibaut and Izacard, Gautier and Martinet, Xavier and Lachaux, Marie-Anne and Lacroix, Timoth{\'e}e and Rozi{\`e}re, Baptiste and Goyal, Naman and Hambro, Eric and Azhar, Faisal and others},
  journal={arXiv preprint arXiv:2302.13971},
  year={2023}
}

@inproceedings{decao2021genre,
  author    = {Nicola {De Cao} and
               Gautier Izacard and
               Sebastian Riedel and
               Fabio Petroni},
  title     = {Autoregressive Entity Retrieval},
  booktitle = {9th International Conference on Learning Representations, {ICLR} 2021,
               Virtual Event, Austria, May 3-7, 2021},
  publisher = {OpenReview.net},
  year      = {2021},
  url       = {https://openreview.net/forum?id=5k8F6UU39V},
}

@inproceedings{tang2023gensurvey,
author = {Tang, Yubao and Zhang, Ruqing and Guo, Jiafeng and de Rijke, Maarten},
booktitle = {SIGIR-AP 2023: 1st International ACM SIGIR Conference on Information Retrieval in the Asia Pacific},
month = {November},
pages = {294--297},
publisher = {ACM},
title = {Recent Advances in Generative Information Retrieval},
year = {2023}}

@inproceedings{feng2022recforest,
 author = {Feng, Chao and Li, Wuchao and Lian, Defu and Liu, Zheng and Chen, Enhong},
 booktitle = {Advances in Neural Information Processing Systems},
 doi = {10.52202/068431-2820},
 editor = {S. Koyejo and S. Mohamed and A. Agarwal and D. Belgrave and K. Cho and A. Oh},
 pages = {38912--38924},
 publisher = {Curran Associates, Inc.},
 title = {Recommender Forest for Efficient Retrieval},
 url = {https://proceedings.neurips.cc/paper_files/paper/2022/file/fe2fe749d329627f161484876630c689-Paper-Conference.pdf},
 volume = {35},
 year = {2022}
}

@inproceedings{he2026plum,
  title={Plum: Adapting pre-trained language models for industrial-scale generative recommendations},
  author={He, Ruining and Heldt, Lukasz and Hong, Lichan and Keshavan, Raghunandan and Mao, Shifan and Mehta, Nikhil and Su, Zhengyang and Tsai, Alicia and Wang, Yueqi and Wang, Shao-Chuan and others},
  booktitle={Proceedings of the ACM Web Conference 2026},
  pages={8093--8104},
  year={2026}
}










\end{document}